\documentclass[lettersize,journal]{IEEEtran}

\IEEEoverridecommandlockouts                              
\ifCLASSINFOpdf
\else
\fi
\usepackage{amsmath}
\usepackage{amsthm}
\usepackage{graphicx}          
\usepackage{amsmath} 
\usepackage{amssymb}  
\usepackage{color,soul}
\usepackage{enumitem}
\usepackage{flushend}
\usepackage{cite}
\usepackage{tabularx}

\allowdisplaybreaks

\newtheorem{remm}{Remark}

\usepackage{arydshln}

\begin{document}

%
\title{Exploring Adversarial Threat Models \\in Cyber Physical Battery Systems}

%
\author{Shanthan Kumar Padisala,~
        Shashank Dhananjay Vyas,~and~Satadru Dey
\thanks{S. K. Padisala, S. D. Vyas, and S. Dey are with the Department of Mechanical Engineering,
        The Pennsylvania State University, University Park, Pennsylvania 16802, USA.{\tt\small \{sfp5587,sbv5192,skd5685\}@psu.edu}.}
}

\maketitle

\begin{abstract}
Technological advancements like the Internet of Things (IoT) have facilitated data exchange across various platforms. This data exchange across various platforms has transformed the traditional battery system into a cyber physical system. Such connectivity makes modern cyber physical battery systems vulnerable to cyber threats where a cyber attacker can manipulate sensing and actuation signals to bring the battery system into an unsafe operating condition. Hence, it is essential to build resilience in modern cyber physical battery systems (CPBS) under cyber attacks. The first step of building such resilience is to analyze potential adversarial behavior, that is, how the adversaries can inject attacks into the battery systems. However, it has been found that in this under-explored area of battery cyber physical security, such an adversarial threat model has not been studied in a systematic manner. In this study, we address this gap and explore adversarial attack generation policies based on optimal control framework. The framework is developed by performing theoretical analysis, which is subsequently supported by evaluation with experimental data generated from a commercial battery cell.
\end{abstract}


%
\IEEEpeerreviewmaketitle

\section{Introduction}

\subsection{Motivation}
Technological advancements like IoT, advanced communication systems, and cloud computing protocols have enabled remote control of batteries, transforming traditional batteries into cyber physical systems having various advantages. Examples include smart distributed control of power grids' Battery Energy Storage Systems (BESS) and cloud-based wireless Battery Management Systems (BMSs) \cite{gm_corporate_newsroom,bosch_mobility_solutions}.

Despite numerous benefits, extensive communication networks make BMS vulnerable to cyber attacks. An adversary can have access to multiple attack surfaces and corrupt the transmitted voltage and current data -- manipulating the power requirements and current drawn. In \cite{bhusal2021cybersecurity}, a review of cyber security aspects of smart charging management systems of Electric Vehicles (EV) is provided. The work in \cite{kharlamova2020cyber} talks about cyber threats in BESSs. A framework to assess the impacts of cyber attacks in EVs is developed in \cite{sripad2017vulnerabilities}. The most common results of these cyber attacks are over-charging leading to thermal runaway and battery explosion and over-discharging leading to insufficient power delivery.

\subsection{Need for Adversarial Threat Models}
However, most of these aforementioned approaches do not focus on understanding the adversarial dynamics. Adversarial dynamics refers to the process followed by the adversary to develop the attacks which can then be injected in the system. In \cite{7408352}, modeling the adversarial dynamics is described in detail. Such modeling efforts are needed to understand the dynamics of the adversarial attacks in the first place which gives us intuition into the decision-making process of the adversaries and then using that information we can develop and deploy technological countermeasures. The modeling process essentially involves designing the attack signals so that the adversarial objectives are satisfied. One of the major objectives of the adversaries is to remain stealthy. Stealthiness means that the attack is not reflected at the user's end even though the attack may have already been injected by the adversary. Understanding the adversarial dynamics helps us greatly in such scenarios to design resilient control strategies.

\subsection{{Literature Review}}

Few works proposed solutions to the cyber physical security of BMSs. {A framework for identifying security concerns in electric vehicles has been presented in \cite{8813669}. In \cite{culler2021cybersecurity}, the impact of attacks on grid-connected batteries on several aspects of grid has been explored. The work in \cite{10630530} developed a dataset for analyzing security aspects of batteries. In \cite{9507536}, a man-in-the-middle type stealthy attacks on battery systems has been presented. The work in \cite{7937798} presented a control scheme for cyber-attacks on distributed energy storage systems. A risk analysis has been performed in \cite{7723824} for microgrids with attacks on solar and energy storage control systems.} In \cite{rahman2018study}, cybersecurity of corrupting State-of-Charge (SOC) data of EV battery's is explored for which the authors trained a Back Propagation Neural Network and tested on experimental dataset to estimate battery SOC under nominal and attacked settings. A Convolutional Neural Network based framework to detect and classify attacked battery sensor data and communication data is proposed in \cite{lee2021convolutional}. The work in \cite{dey2020cybersecurity} attempts to address the issue of cybersecurity of plug-in EV using mathematical conditions of stealthy cyber attacks and developing model-based attack detection schemes. 

There exist some efforts in literature towards threat modeling of various cyber-physical systems. For example, the work \cite{zografopoulos2021cyber} explored threat models of cyber-physical energy systems. In \cite{bhattacharya2020automated}, a reinforcement learning based approach was presented for automated adversarial emulation in general cyber-physical systems. The work \cite{khan2017stride} also explored a systematic approach for threat modeling in cyber-physical processes. However, most of these existing works either focus on general cyber-physical system without consideration of specific application-focused dynamics or deal with applications different from battery systems. In terms of battery systems, the work in \cite{dey2020cybersecurity} briefly discussed stealthy attack generation which captures the relationship between input and output attacks for stealthiness. Similarly, the work in \cite{Srinath} utilizes an adversarial threat model that considers overcharge/discharge objectives in designing input and output attacks. However, this adversarial model is formulated in an ad-hoc manner -- without applying a systematic approach. Specifically, in order to understand the potential cyber threats to battery systems, an adversarial threat model has not been systemically studied which considers all of the following aspects: (i) how the primary purpose of the battery system is hampered, that is, how the power delivering and/or charging capabilities can be potentially compromised with specific over-charging and over-discharging objectives; (ii) how battery safety is adversely affected under attacks; (iii) how stealthiness of attacks can be maintained; and (iv) how physical knowledge of the battery systems (current-voltage-charge dynamics), along with multi-objective adversarial goals arising from unsafe state generation and stealthiness, can potentially be leveraged to inject the aforementioned attacks. This work addresses these gaps by developing an adversarial modeling framework which formulates an multi-objective optimal control problem for potentially undesirable/unsafe operations and utilizes a combination of model and feedback data to enforce stealthiness.

{There is a major challenge in developing a systematic framework for adversarial threat modeling in cyber-physical battery systems. Modeling adversarial behavior can be pursued in a manner similar to behavioral modeling framework. However, such behavioral models are often based on data-driven approaches requiring extensive amount of training and testing data -- which are not available for the battery application. However, it should also be noted that the a major subset of adversarial objectives aims to create certain physical disruption. That is, the adversary will potentially use the knowledge of battery's physical behavior to create such disruption. For example, the objective of overcharging or undercharging a battery would require the understanding of how applied battery current affects battery state-of-charge. These physically disruptive attacks can be analyzed by focusing on the battery systems physical behavior which gives us insights about potentially harmful input current sequences. Furthermore, stealthiness requires that the battery output signals should behave nominally even under harmful input current attacks. This also needs an understanding of battery physical behavior such that the effect of harmful current is masked at the battery output signals. In other words, the physical dynamics of the battery systems can be exploited to gain a better understanding of adversarial behavior. While such understanding can be formulated in a theoretical framework, there are limitations in such approach which arise from the potential inaccuracies of battery models. Such limitations can be addressed by using experimental data from commercial battery cells along with the mathematical models -- which can potentially suppress the battery model uncertainties to certain extent. Following these arguments, in the proposed framework, we leverage the physical battery dynamics along with experimental battery data to develop an understanding of potential adversarial behaviors.}

\subsection{{Contribution}}
{In light of the aforementioned research needs and gaps, the main contribution of this work is as follows: This paper explores adversarial threat models in CPBS by analyzing cyber attack generation policies with high-level adversarial objectives and stealthiness features. The approach is formulated based on optimal control principles. Theoretical analysis is used to develop the framework. Subsequently, experimental data generated from a commercial battery cell is used to evaluate the effectiveness under real-world scenarios. We re-iterate the main differences between \cite{dey2020cybersecurity,Srinath} and this paper: (i) \cite{dey2020cybersecurity} and \cite{Srinath} focus on attack detection and detection-mitigation, respectively, whereas this current work focuses on understanding and formulating attack injection mechanisms in detail; (ii) current work utilizes a systematic optimal control approach to generate the current attacks, whereas \cite{dey2020cybersecurity} and \cite{Srinath} used an ad-hoc approach; and (iii) current work utilizes an experimental data and experimentally identified battery model capturing real-world scenarios, whereas \cite{dey2020cybersecurity} and \cite{Srinath} used simulated battery data.} 

The rest of the paper is organized as follows. Section II states the problem, Section III describes the methodology for attack generation policies, Section IV discusses results, and Section V concludes the paper.

\section{Problem Statement: Adversarial attacks in Battery Systems}
\begin{figure}[b]
    \centering
    \vspace{-3mm}
    \includegraphics[scale = 0.33]{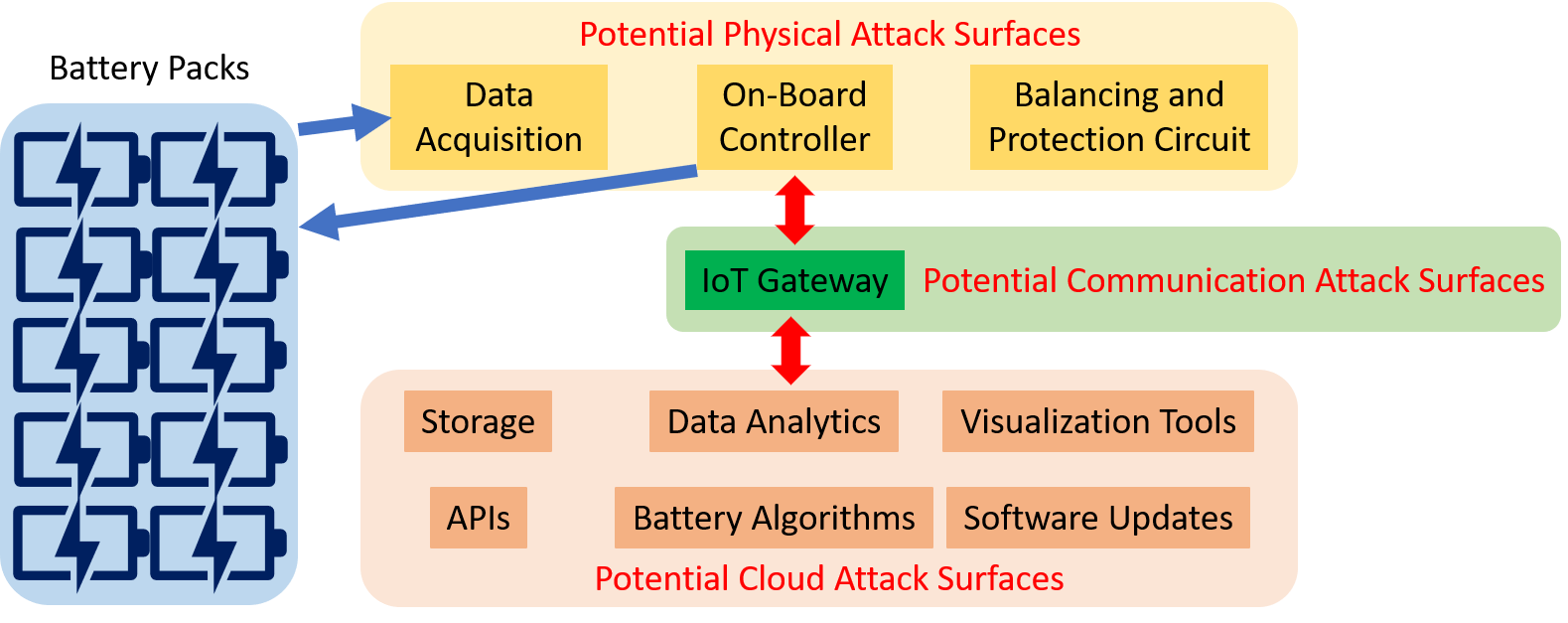}
    \vspace{-5mm}
    \caption{A simplified schematic of cyber physical BESS, adopted and modified from \cite{kim2018cloud}.}
    \label{fig:AttackSurfaces}
\end{figure}

A simplified architecture of a cyber physical BESS system is shown in Fig. \ref{fig:AttackSurfaces}. As compared to traditional battery systems which have only one access point, a cyber physical system can be said to have three access points: (1) Physical access (sensors, onboard controllers, and diagnostic modules, etc.); (2) Communication networks (LTE, 5G, Bluetooth, WiFi, etc and IoT devices); (3) Cloud services (data storage centers, services offered through cloud - like data analytics, visualization tools, firmware updates, etc). All these access points are interconnected as is evident from Fig. \ref{fig:AttackSurfaces}. Since the attack surface increases due to the interconnections, the CPBS becomes more prone to adversarial attacks. 

{When the traditional battery systems are converted into CPBS, their functionality is enhanced due to all the additional software and hardware integrations that involve sensing, computation and communication technologies. However, these enhancements come with a downside which compromises the operational safety of the CBPS by introducing vulnerabilities. These vulnerabilities can potentially range from issues arising in sensing and actuation to compromised monitoring to software and firmware updates leading to malfunctions. For example, False Data Injection Attacks (FDIA) involve manipulation of the sensors’ and actuators’ data causing the BMS algorithms to receive incorrect feedback signals. These types of the attacks can be termed as man-in-middle attacks \cite{kharlamova2020cyber}. These vulnerabilities at sensors' and actuators’ ends act as the points of attacks where an adversary can alter the system performance through attack inputs. Eventually, such degraded system performance can create an unsafe operating conditions for the CPBS.}

Due to the presence of WiFi or Bluetooth connectivity, the adversary can eavesdrop into this network from a short distance. The common entry points, devices through which an attacker remotely connect with the system, are service equipment, public-facing infrastructure, vendor cloud service or server, Local Area Network (LAN), WiFi or bluetooth connected devices, meter, software and firmware upgrades, Virtual Private Network (VPN), Virtual Network Computing, etc. \cite{trevizan2022cyberphysical}. Given the large amounts of security risks faced by CPBS, it is essential to understand the adversarial nature of attackers and hence we focus on the cyber attack perspective. Particularly, we formulate our problem from the point of view of adversaries creating stealthy cyber attacks. This will help us in designing defense strategies against cyber attacks using the knowledge of adversaries obtained by generating stealthy attack policies. {The integration of IoT into CPBS introduces various potential attack vectors as this integration adds network connectivity and remote access to physical systems, thereby making them vulnerable to cyber-attacks \cite{Kumbhar2018}. These IoT devices in CPBS often collect real-time data allowing automation or remote control of battery operations, which increases the system's attack surface. As mentioned above, there are various types of attacks that can occur – remote access exploit, data manipulation, denial of service attack, etc. And all these attacks at the end lead to manipulation of the system at sensors and actuators level. Sensors like voltage, current, temperature sensors are the primary sensors that are most susceptible to manipulation as these are the sensors that rely on direct physical measurements and can be spoofed or altered through attacks. This is the main working principle of one of the FDIA attacks – man-in-middle attack.}

{Given that the transformation of battery systems into cyber-physical systems is a relatively new and emerging area, the literature on cyber-physical security in such energy storage systems is limited -- especially on understanding the attack generation and injections. The work in Kharlamova et al. \cite{KHARLAMOVA2023107795} mentions an attack against a particular battery service where an attack is aimed at preventing the control unit from having a realistic battery SoC. In this work, different attack vectors are injected on all the measurements thereby corrupting the measurements. In \cite{10230848}, false data injection attacks on input current have been explored. Similarly, false data injection attacks on voltage sensors are discussed in \cite{9744036}. Similar mentions of these types of attacks were made in \cite{kharlamova2020cyber} and \cite{dey2020cybersecurity}.}

For the current work, the battery system is assumed to be just limited to the current as the only input and voltage measurement as the only output. In such a type of system, the only way for a cyber attacker to attack this CPBS is by introducing attacks in the form of an additional current. 
The choice of this additional current injection will cause the system to operate in an undesirable unsafe operating region that can be hazardous. In order to introduce the attack in a stealthy fashion, the obvious effects that will be seen in the output voltage measurements must be mitigated so that the outputs will be as close as to the originally anticipated values. This way, the attack injected will be stealthy and will be undetected. The imperfections in the output measurements will be misunderstood to the measurement noises or any other modeling uncertainties by the user. In short, the main objective of the current work is to propose one of the only few potential ways an attacker can introduce a cyber attack into the CPBS and yet stay stealthy. Leveraging this, one can design potential ways to detect and mitigate such cyber attacks that can be hazardous. 

{Note that there are certain protection protocols in existing BMSs. For example, over-current protection typically uses a physical mechanism (e.g., an electronic switch/fuse) to disconnect the battery when the charging/discharging current exceeds a pre-defined threshold. However, the attacks can still be implemented while the current is well within such pre-defined current limits. That is, injecting a safe current for a longer time can also lead to overcharging or discharging without triggering the over-current protection mechanism. Other than over-current, there is also over-voltage and over-temperature protection -- which are typically done by checking if the measured terminal voltage and temperature have crossed pre-defined limits. However, such limit checking is typically performed within the BMS software-based computation. Such voltage and temperature overprotection can be overridden either by manipulating the measurement reading entering the computation or by changing the limits within the computation. Other than using the over-voltage or under-voltage condition checks, there can be overcharging or overdischarging protection which utilize the state-of-charge information. Note that such state-of-charge is also computed within the BMS where the attacker can potentially change the current information entering the BMS or the computation itself in order to bypass the overcharge/overdischarge protection. Along similar lines, the work in \cite{culler2021cybersecurity} showed that by changing some computation/model (e.g., the volt-VAR curve), battery set-point can be manipulated -- leading to potentially destabilizing effects.}

\begin{figure}[h!]
    \centering
    \vspace{-3mm}
    \includegraphics[scale = 0.28]{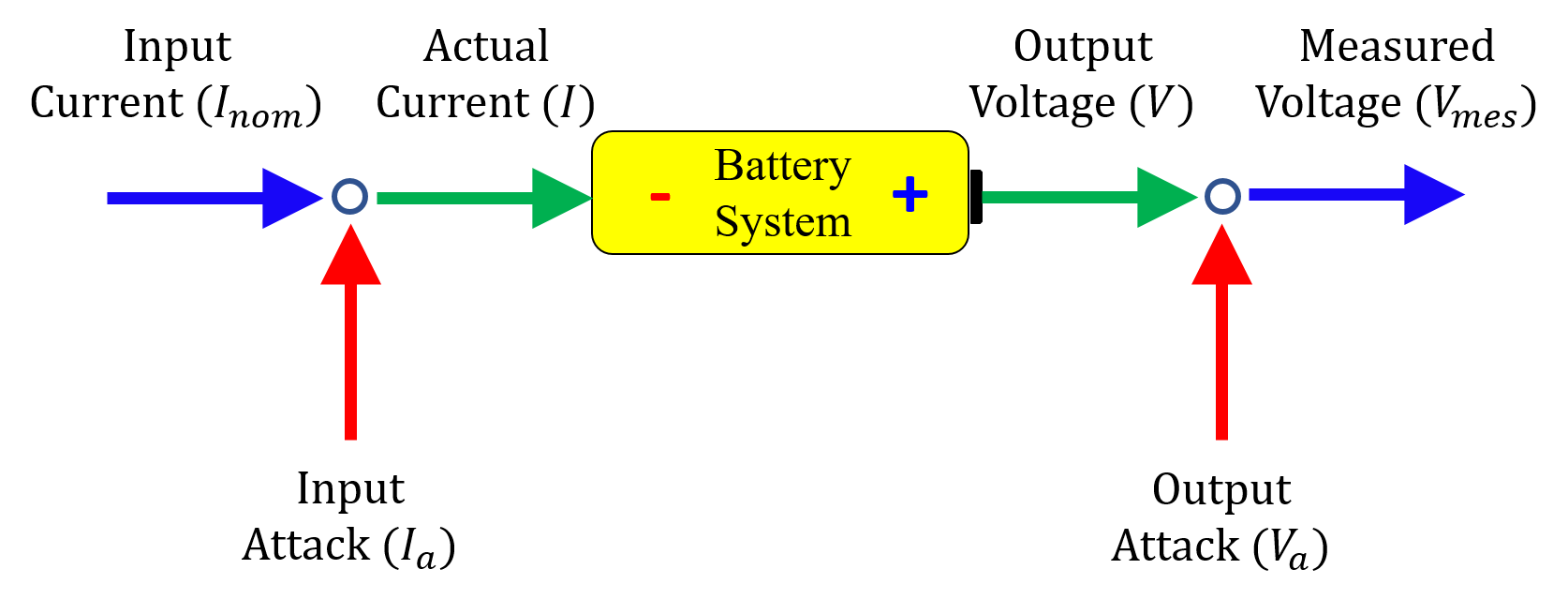}
    \vspace{-5mm}
    \caption{Adversarial attack injection into battery systems.}
    \label{fig:Adverserial_Attack_Strategy}
\end{figure}

A schematic of the potential attack by the cyber attacker in the system of CPBS is illustrated in Fig. \ref{fig:Adverserial_Attack_Strategy}. It can be seen that the original intended current input to the battery cell is depicted by $I_{nom}$, which however is altered into an actual input $I$ due to the additional attack current $I_a$, which is introduced by the cyberattacker. This results in an output voltage of $V$, however again due to the addition of voltage attack $V_a$ on the outputs end, the voltage as seen by the battery management systems (BMS) is going to be $V_{mes}$.

\begin{remm}\normalfont
{In this problem formulation, we have considered the input to be the applied battery current and the output to be the measured terminal voltage. Note that in typical commercial battery systems, there are two other possible input and output: cooling system and measured temperature. However, both of the cooling and temperature have much slower dynamics than the considered electrical quantities (current and voltage). In order to inject attacks that are constrained to achieve adversarial objectives faster, electrical quantities can potentially be the primary choice given their faster response time. In other words, the effects of attack will show up faster in the electrical quantities as compared to the thermal quantities. Due to this reason, we focus on input current and output voltage scenario in this work. Furthermore, even scaling the problem formulation from cell to pack level, electrical quantities will still represent the faster set of variables -- with the modification that cell current and voltage will be replaced with pack-level available currents and voltages, respectively. Finally, the proposed optimal control formulation is generic enough to be extended to include cooling actuation and temperature measurements with minor modifications in the solution and including of thermal dynamics along with electrical dynamics. We consider such extension to be a part of future work.}
\end{remm}



\section{Adversarial threat model}
{In this section, we discuss the attack generation mechanisms in detail. First, we describe the battery equivalent circuit model (ECM) which is the basis for model-based attack generation mechanisms. Then, we focus on the optimal control based input attack generation. Finally, we discuss the output attack generation based on open-loop model and voltage feedback.}

\subsection{Description of battery equivalent circuit model (ECM)}
\label{Section3a}
\begin{figure}[h!]
   \centering
   \includegraphics[scale = 0.35]{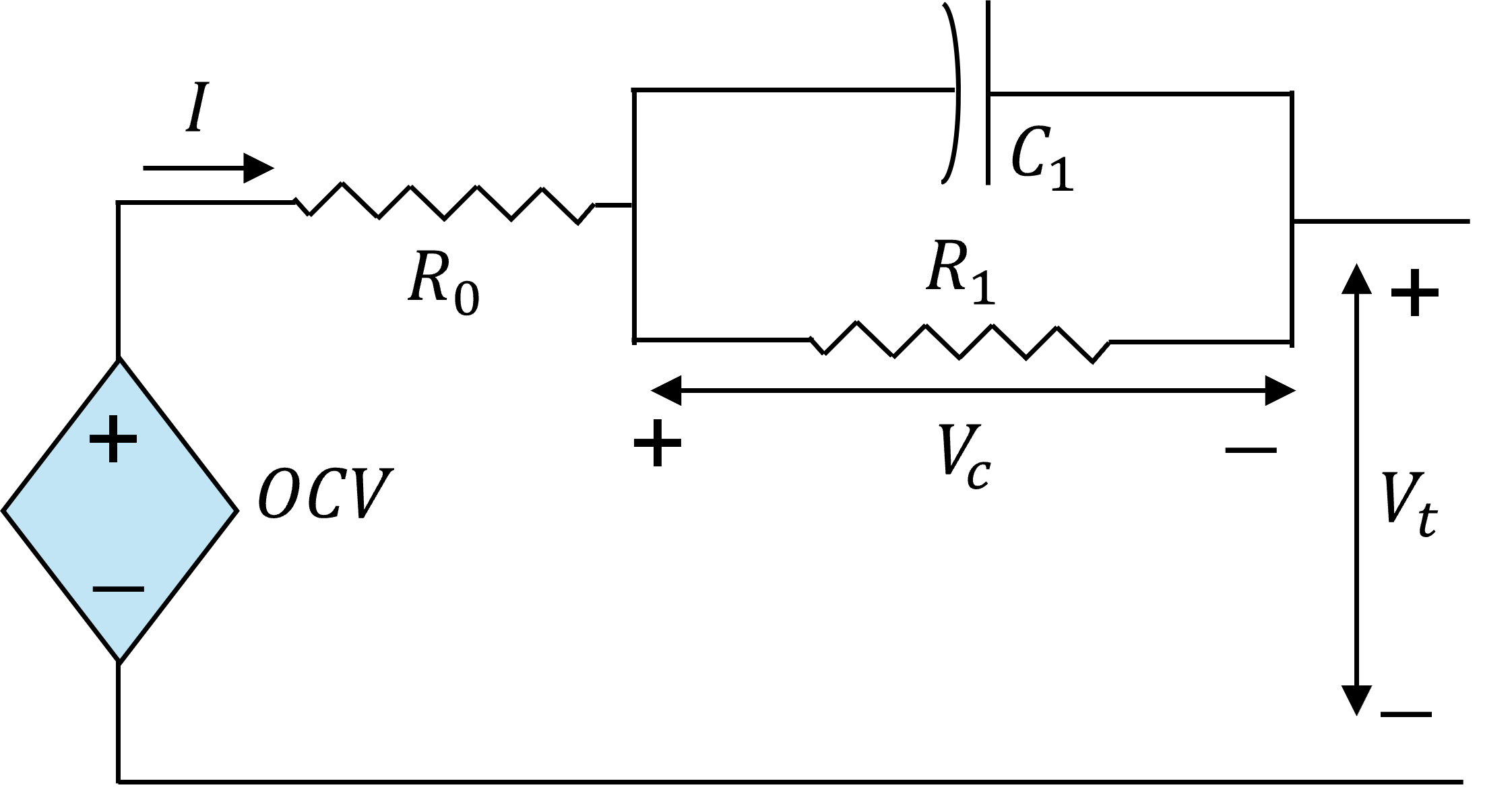}
   \caption{{First order equivalent circuit battery model.}}
   \label{fig:ECM}
\end{figure}

A first order ECM \cite{barsoukov1999universal} is used to model the battery (shown in Fig. \ref{fig:ECM}). This includes a set of resistor and capacitor in parallel connected in series with another resistor. This ECM system can be described by a system of equations as given below:
\begin{align}
    & \dot{SoC} = -\frac{I}{Q}, 
    \dot{V_c} = -\frac{1}{R_1C_1}V_c + \frac{I}{C_1},\\
    & V_t = OCV(SoC) - V_c - IR_0,\label{eqn:V_t}
\end{align}
{where $Q$ is the battery capacity in $[A-s]$; $R_0$ is the series resistance, $R_1$ is the resistance parallel to the capacitance -- both in $[\Omega]$; $C_1$ is the ECM capacitance in $[F]$; $V_c$ is the voltage across the capacitor, $V_t$ is the terminal voltage, and $OCV$ in the open circuit voltage -- all of them in $[V]$, and $SoC$ is the battery state of charge (dimension-less variable in [0,1]).}

\begin{remm}\normalfont
{Note that the ECM mentioned above is based on certain assumptions. First, ECM provides a lumped modeling approach without consideration of battery system's internal electrochemical-thermal-mechanical phenomena. Next, the parameters of the ECM are considered to be fairly constant. Finally, only the first order dynamics have been incorporated ignoring higher order dynamical aspects. These assumptions enabled us to have a computationally efficient model with desired mathematical structure -- aiding our subsequent analysis.}
\end{remm}

\begin{remm}\normalfont
{The adversary is mainly concerned ``input-output'' and ``state-of-charge'' behavior of the battery. This is because the adversary will manipulate only the ``input'' current, and suppress the effect on the ``output'' voltage – and will intend to cause undesirable ``state-of-charge'' dynamics. The information is sufficient to achieve adversarial objectives -- without the need to know/identify underlying complex electrochemical phenomena. Hence, we have chosen a simplified equivalent circuit model for this approach.}
\end{remm}

\subsection{State-space representation of battery systems}
Next, we formulate the ECM equations to a state-space format as given by:
\begin{align}
    &\dot{X} = AX + BU, Y = g(X,U), 
    \label{eqn:output}
\end{align}
where $X = \begin{bmatrix} SoC & V_c \end{bmatrix}^T$ is the state vector, $U=I$ is the current input, and $Y=V_t$ is the output terminal voltage. The function $g(.)$ is formed by the output voltage equation and the $A$ and $B$ matrices are given by:
\begin{align}
    \label{eqn:statespacematrices}
    & A = \begin{bmatrix} 0 & 0 \\ 0 & \frac{-1}{R_1C_1} \end{bmatrix}, \ B = \begin{bmatrix} \frac{-1}{Q} \\ \frac{1}{C_1}\end{bmatrix}.
\end{align}

In the presence of cyber attacks, we modify \eqref{eqn:output} as
\begin{align}
    &\dot{X} = AX + B(U_{nom}+U_a), \label{eqn:ActualStateSpace-attack}\\
    & Y = g(X,U_{nom}+U_a)+Y_a, \label{eqn:output-attack}
\end{align}
where the input $U$ applied to the battery system has two components: $U_{nom}$ is the nominal input applied by the user and $U_a$ is the additional input current injected by the attacker. Furthermore, $Y_a$ is the output attack injected by the attacker to corrupt the measured voltage $Y$. Next, we discuss potential ways to design the attack signals $U_a$ and $Y_a$. Specifically, the design of $U_a$ will be based on the objective of causing adversarial disturbances. We will consider two cases of adversarial disturbances: (i) \emph{overcharging:} Here, the adversary will aim to overcharge the battery system with a higher charge value compared to the user's desirable charge reference. (ii) \emph{over-discharging:} Here, the adversary will aim to over-discharge the battery system with a lower discharge value compared to the user's desirable discharge reference. Subsequent to the design of $U_a$, the design of $Y_a$ will be based on the objective of adversarial stealthiness. In the next two subsections, we will discuss the designs of $U_a$ and $Y_a$.

\subsection{Optimal control-based input attack generation}
\label{InputAttackGeneration}

In this subsection, we will discuss the potential design of input attack signals, denoted by $U_a$ in \eqref{eqn:ActualStateSpace-attack}-\eqref{eqn:output-attack}. The $U_a$ is essentially an additive current signal that is superimposed with user's intended current signal $U_{nom}$. To this end, the adversary's objective is to cause either overcharging or over-discharging. Both of these objectives can be formulated as following a reference state trajectory desired by the adversary. This reference trajectory will enable the adversarial requirements of overcharging and over-discharging. The adversarial objective has two components: (i) induce unsafe states of overcharging or over-discharging as fast as possible by applying input attack $U_a$, and (ii) achieve such unsafe state with minimum possible efforts, that is, using limited $U_a$ signal energy. 

The adversarial process behind attack generation has two components. First, the attack should hamper battery operation by either diminishing its power delivering capability by over-discharging or by creating unsafe situation by over-charging. This can be done if the battery charge follows a prescribed charge trajectory that explicitly over-charges or over-discharges to a desired point. Minimizing the difference between the prescribed trajectory and actual battery charge at all times will ensure satisfaction of this first component. As for the second component, the attack should use the battery energy in an intelligent manner so that hardware current limits are not hit while ensuring that the attack does not require a prolonged injection time. Minimizing the required attack energy will satisfy the second component. Now, these two components need to be put into same mathematical abstraction in order to find attacks that satisfy both. Optimal control theory provides an unique theoretical abstraction where (i) multiple adversarial objectives can be unified (for example, as done in \cite{chen2020optimal}), and (ii) efficient techniques exist to find the optimal solution \cite{zhu2018optimal,lessard2019optimal}. Hence, optimal control framework can effectively model and analyze adversarial attack generation policies for battery systems. Subsequently, such optimal control based attack model will help build resilience by: (i) creating attack detection strategies leveraging the optimal threat model, (ii) securing the data transmission among the vulnerable parts of the battery, and (iii) providing insights into developing hardware-based solutions to combat cyber threats in batteries.

{As mentioned before, both of the adversarial objectives can essentially be formulated as optimization problems. For example, the first objective can be expressed into two parts: the first part corresponds to following a reference state trajectory (that is, an $SoC$ trajectory over time) as closely as possible throughout the attack action time whereas the second part corresponds to achieving a desired final state at the end action time as closely as possible (that is, the final $SoC$ desired by the adversary). Following these desired state reference trajectories as closely as possible can be incorporated as minimization of the corresponding tracking errors -- that is, minimization of the terms $\int_0^{t_f} \left\|(X(t)-X_{ref}(t))\right\|^2 dt$ and $\left|X(t_f)-X_{ref}(t_f)\right|^2$ with $X(t)$ and $X_{ref}$ being the actual and desired states, and the attack occurs in $[0,{t_f}]$ time-period. The second objective can also be expressed as the minimization as of the attack current energy throughout the attack action time -- that is, minimization of the term $\int_0^{t_f} \left|U_a(t)\right|^2 dt$. Furthermore, these two objectives sometimes can be in conflict with each other. For example, a higher value of $U_a$ will induce fast overcharge/over-discharge but at the cost of higher attack energy requirement. On the other hand, a lower value of $U_a$ will eventually induce the attack but at a prolonged time. This trade-off can be incorporated as the the weighted sum of the aforementioned minimization terms, where the weights can be chosen to control the trade-off arising due to the potential conflict. In summary, we see that the adversarial objectives can be framed into a sum of weighted terms which should be minimized. In other words, an optimization approach is deemed suitable to formulate the input attack design problem. Specifically, given the nature of the real-time nature of attack injection, an optimal control approach that considers these two potentially conflicting components along with the dynamical nature of battery systems will capture the essence of the attack generation mechanism. It should be noted that such optimal control approach requires the assumption that the adversary has a reasonably accurate model of the battery system under consideration. That is, certain level of model knowledge is required to have a closed-form analytical model of adversarial behavior in terms of current attack injection protocol.}

Following the above argument, we resort to an optimal control problem with the following potential adversarial objective function:
\begin{align}
    \min_{U_a} J & = \frac{1}{2}[X(t_f)-X_{ref}(t_f)]^TQ_1[X(T)-X_{ref}(T)] \nonumber\\
    & + \frac{1}{2} \int_{t_0}^{t_f}[X(t)-X_{ref}(t)]^TQ_2[X(t)-X_{ref}(t)] dt \nonumber\\
    &+ \frac{1}{2} \int_{t_0}^{t_f} R U_a^2(t) dt, \label{eqn:ObjectiveFxn}
\end{align}
where the time window of attack is $t \in [t_0, t_f]$, $X_{ref}$ is a (potentially) time-varying state reference trajectory aiming to either overcharge or over-discharge the battery. The first and second terms on the right hand side of \eqref{eqn:ObjectiveFxn} ensure that the actual state $X$ follow the reference $X_{ref}$ as closely as possible during the time period and in the final time $t_f$; and the matrices $Q_1$ and $Q_2$ are adversary-defined weighting matrices. The third term on the right hand side of \eqref{eqn:ObjectiveFxn} limits the magnitude of current attack signal $U_a$ with a weighting factor $R$. To this end, the goal of the adversary is to minimize the objective function \eqref{eqn:ObjectiveFxn} with respect to the dynamic constraints given as system model \eqref{eqn:ActualStateSpace-attack}-\eqref{eqn:output-attack}.

In order to solve the aforementioned optimal control problem, we resort to the  the Lagrange multiplier and sweep method discussed in \cite{lewis2012optimal,bryson1975applied}. First, considering the objective function \eqref{eqn:ObjectiveFxn} and the state dynamics constraint \eqref{eqn:ActualStateSpace-attack}, we construct the Hamiltonian as follows \cite{lewis2012optimal,bryson1975applied}:
\begin{align}
    H(t) &= \frac{1}{2}[X(t)-X_{ref}(t)]^TQ_2[X(t)-X_{ref}(t)] \nonumber\\
    &+\frac{1}{2}\lambda^T[AX(t)+BU_a(t)+BU_{nom}(t)],    \label{eqn:Hamiltonian}
\end{align}
where $\lambda$ is the Lagrange multiplier. Subsequently, the co-state equations are given by:
    \begin{align}
        &\frac{\partial H}{\partial U_a} = 0 \implies RU_a + B^T\lambda = 0 \implies U_a = -R^{-1}B^T\lambda, \label{eqn:CS1}\\
        &\frac{\partial H}{\partial X} = -\dot{\lambda} \implies  Q_2[X-X_{ref}] + A^T\lambda = -\dot{\lambda}, \label{eqn:CS2}
    \end{align}
To solve these two co-state equations \eqref{eqn:CS1} and \eqref{eqn:CS2}, $\lambda$ is chosen of the form $\lambda = SX-V$, where $S$ and $V$ are variables to be determined \cite{lewis2012optimal,bryson1975applied}. Upon taking the derivative of this equation with respect to time, we get the following form:
\begin{align}
    &\dot{\lambda} = \dot{S}X + S\dot{X} - \dot{V} \nonumber \\
    \implies &\dot{\lambda} = \dot{S}X + S[AX+BU_a+BU_{nom}] - \dot{V} 
\end{align}
Upon substituting the expression of $\dot{\lambda}$ and ${\lambda}$ into \eqref{eqn:CS2}, we get
\begin{align}
        & Q_2[X-X_{ref}] + A^T[SX-V] \nonumber \\
        & = -\dot{S}X - S[AX-BR^{-1}B^T(SX-V)+BU_{nom}] + \dot{V}. \label{eqn:CS2new}
    \end{align}
The equation \eqref{eqn:CS2new} can be decoupled as two Riccati equations:
\begin{align}
    \dot{V} + A^TV - SBR^{-1}B^TV - SBU_{nom} + QX_{ref} &= 0, \label{eqn:RiccattiEquations1} \\
    \dot{S} + SA - SBR^{-1}B^TS + A^TS + Q &= 0 \label{eqn:RiccattiEquations2}
\end{align}

The boundary conditions are computed to be:
\begin{align}
    S(T) = Q_1, V(T) = Q_1X_{ref}(T) \label{eqn:BC2}
\end{align}

Solving \eqref{eqn:RiccattiEquations1}-\eqref{eqn:RiccattiEquations2} will give us $S$ and $V$, and subsequently, we can compute $U_a$ as $U_a = -R^{-1}B^T[SX-V]$.

\subsection{Open-loop model and real-time feedback-based output attack generation}
\label{Subsection:OutputAttack}
Once the input current attack is generated, the next focus is on achieving stealthiness. That is, the adversary's goal is not to trigger any alarm when the input attack is injected. To achieve such stealthiness, the adversary designs an output attack signal $Y_a$ which corrupts the measured terminal voltage, as shown in \eqref{eqn:output-attack}. This attack signal $Y_a$ is generated in a way such that the effect of $U_a$ is cancelled by $Y_a$ at the terminal voltage output.

For this output attack generation, the adversary utilizes the knowledge of system model \eqref{eqn:ActualStateSpace-attack}-\eqref{eqn:output-attack}. To this end, $Y_a$ is generated such that the following condition is satisfied:
\begin{align}
    & Y = Y_{nom}, \nonumber\\
    \implies & g(X,U_{nom}+U_a)+Y_a = g(X_{nom},U_{nom}),\nonumber\\
    \implies & Y_a = g(X_{nom},U_{nom})-g(X,U_{nom}+U_a), \label{output_attack1} 
\end{align}
where $X$ is generated by solving $\dot{X} = AX + B(U_{nom}+U_a)$, $X_{nom}$ is generated by solving $\dot{X}_{nom} = AX_{nom} + BU_{nom}$, and $Y_{nom}=g(X_{nom},U_{nom})$ is the output under no attack condition. The condition \eqref{output_attack1} ensures that the measured output $Y$ under the attack should resemble to the output $Y_{nom}$ under user-applied nominal input $U_{nom}$.

However, finding the output attack $Y_a$ using only the open-loop model \eqref{eqn:ActualStateSpace-attack}-\eqref{eqn:output-attack} has a limitation. That is, if the model knowledge is inaccurate (as dictated by the ECM model parameters), the output attack will not be able to cancel the effect of input attack. To remedy such limitation, the adversary can potentially use the real-time feedback of $Y$ to correct the inaccuracies of open-loop model generated output attack. Accordingly, the output attack generation will be modified as:
\begin{align}
    Y_a = g(X_{nom},U_{nom})& -g(X,U_{nom}+U_a) \nonumber\\
    & + K_a(Y-g(X_{nom},U_{nom})),\label{eqn:output_attack2} 
\end{align}
where $Y$ is the measured real-time terminal voltage, and the added second term on the right hand side is the feedback term that corrects the inaccuracies in the open-loop model generated attack signal. The parameter $K_a$ is a feedback gain determined to achieve acceptable amount of correction.



\section{Results and Discussion}
\begin{figure}[h!]
    \centering
    \includegraphics[scale = 0.05]{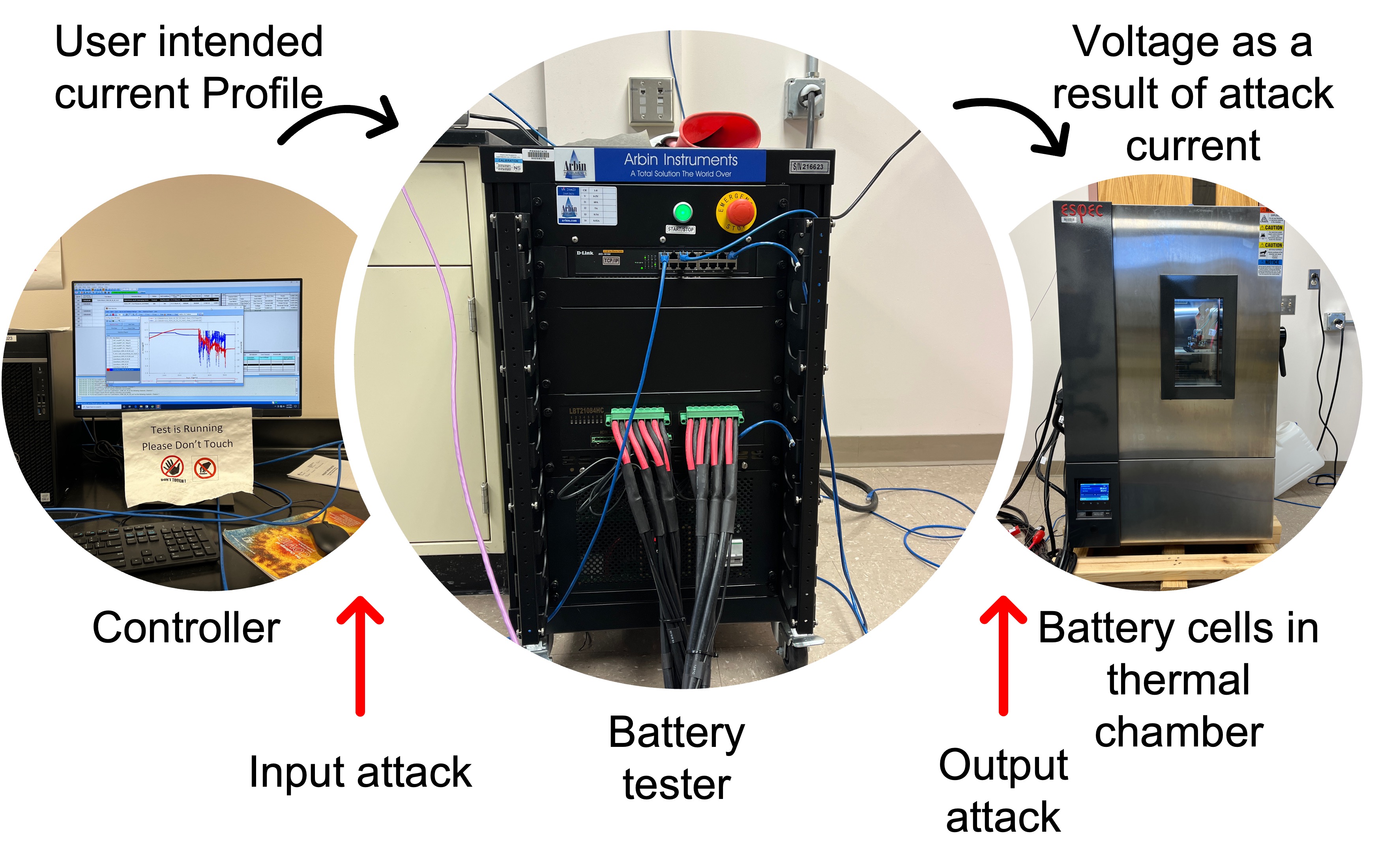}
    \vspace{-5mm}
    \caption{Experimental setup.}
    \label{fig:ExpSetup}
\end{figure}

In this section, we evaluate the proposed threat model using experimental data. First, we discuss the identification of the battery model given in Section \ref{Section3a} from the experimental data. Subsequently, we generate the attack signals described in Section \ref{InputAttackGeneration} and evaluate their efficacy. {Experiments were conducted at 25 $^oC$ temperature in a controlled environment using ESPEC thermal chamber. The input current profile and the output voltage are given and measured by the ARBIN Battery Cyclers. The experimental schedules and the data logging rates were defined in MITS Pro. {MITS Pro is the software which is used to control the battery input current commands, measure temperatures using thermocouples and also to measure the voltage across the battery cell. In other words, in the context of the CPBS this acts as the BMS controller interface level which is responsible for all the tasks like controlling input commands, measuring ouputs, and monitoring the system through temperature feedback. The inputs to the battery cell are given to the MITS Pro software in a file called schedule which has a series of actions to be taken at different times. In the case of an attack, these original intended actions are altered to emulate the attack injection.} An illustration of the experimental setup is depicted in Fig. \ref{fig:ExpSetup}.}

{As mentioned above, in our current experimental setup, the input current control commands to the battery cells are given using MITS Pro software which controls the Arbin Battery Tester. In order to test our testcase of FDIA man-in-middle attack scenario \cite{KHARLAMOVA2023107795} where it is assumed that the attacker gets access to the controllers and tampers with the inputs commands, the fault is injected to the Arbin Battery Tester input profiles. In short, this is basically a result of tampering with the current input commands directly on the Arbin tester. As of now, these profiles are generated offline and injected into the system as this is a preliminary study. However, it is intended to expand this study in the future to an online generation as well as addition of the attacks using DSpace controllers which will emulate the attacker and bypass the actual user intended inputs that are pre-saved at Arbin’s end.}

\begin{figure}[h!]
    \centering
    \vspace{-2mm}
    \includegraphics[scale = 0.5]{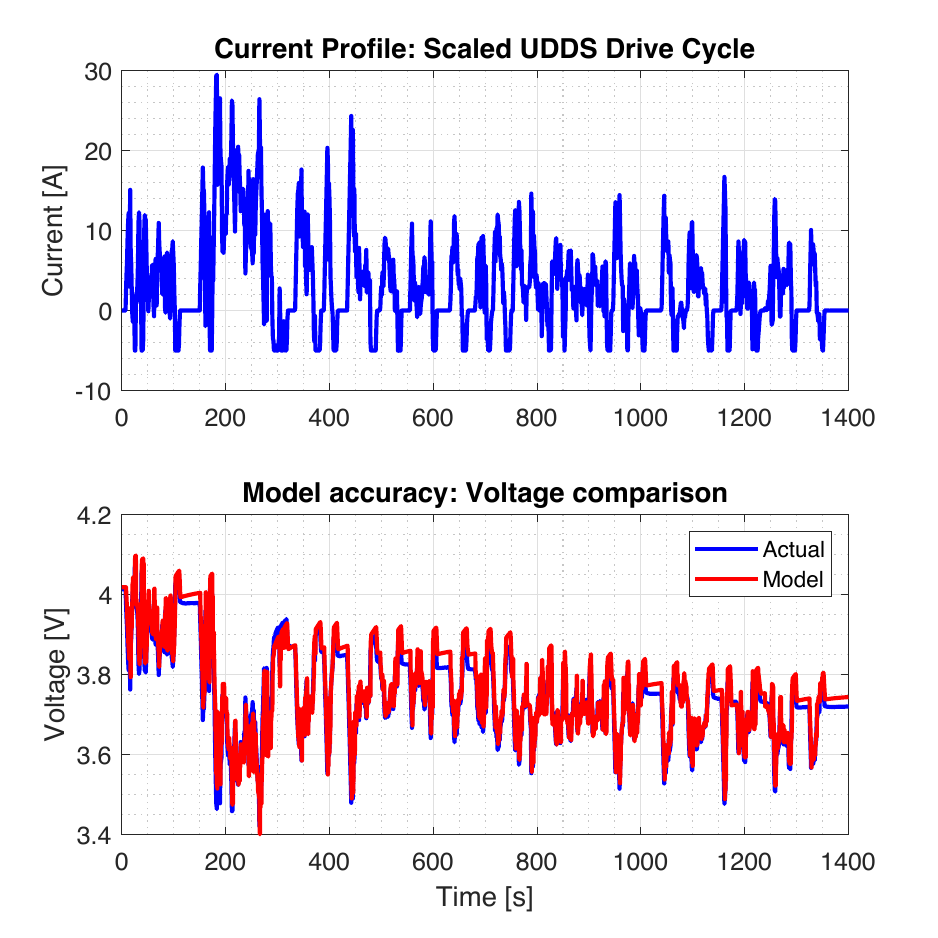}
    \vspace{-5mm}
    \caption{Dynamic current profile and comparison of experimental and model output voltages (RMSE 0.0483 V).}
    \label{fig:ECM_Modelling_UDDS_80_50}
\end{figure}

For this study, we considered a lithium-ion battery cell with 4.2V to 2.5V operating voltages and a rated capacity of 4000 mAh. To identify the battery model parameters, first, the open circuit voltage ($OCV$) of the cell has been experimentally investigated by charging and discharging the battery cell with a very low current. {OCV plays a key role in determining the SoC and State of Health of the cells. OCV is used in BMS to calibrate SOC estimations and make informed decisions regarding the battery operation.} Subsequently, the equivalent circuit parameters $Q$, $R_0$, $R_1$ and $C_1$ are calculated using the Reference Performance Test (RPT) and the identified parameter values are as follows: (1) $Q = 1.4322 \times 10^{4} [A-s]$; (2) $R_0 = 1.3513 \times 10^{-2}[\Omega]$; (3) $R_1 = 1.028 \times 10^{-2} [\Omega]$; (4) $C_1 = 5.2584 \times 10^{3} [F]$. The periodic assessment of battery degradation during life testing is done through RPT - which produces capacity fade, power fade, and impedance rise as a function of test time. After which these parameters are tuned further using a dynamic current profile, generated from the Urban Dynamometer Driving Schedule (UDDS) velocity profile, as shown in Fig. \ref{fig:ECM_Modelling_UDDS_80_50} (top part). A comparison of the experimental voltage and model-generated voltage is shown in Fig. \ref{fig:ECM_Modelling_UDDS_80_50} (bottom part).

To illustrate the effectiveness of the proposed threat model or attack generation approach, we will use the following test cases where the current profiles are derived from UDDS and US06 velocity profiles: 
\begin{itemize}
        \item \textbf{Test Case 1:} Modified UDDS-based profile where user-intended discharging is 80\% to 50\% $SoC$. The adversary's intent is to over-discharge to 20\% $SoC$.
        \item \textbf{Test Case 2:} Modified US06-based profile where user-intended discharging is 80\% to 50\% $SoC$. The adversary's intent is to over-discharge to 20\% $SoC$.
        \item \textbf{Test Case 3:} Modified UDDS-based profile where user-intended charging is 20\% to 50\% $SoC$. The adversary's intent is to overcharge to 80\% $SoC$.
        \item \textbf{Test Case 4:} Modified US06-based profile where user-intended charging is 20\% to 50\% $SoC$. The adversary's intent is to overcharge to 80\% $SoC$.
\end{itemize}

In these aforementioned test cases, the input current attacks are generated using the optimal-control strategy described in Section \ref{InputAttackGeneration}. Subsequently, the output voltage attacks are generated as discussed in Section \ref{Subsection:OutputAttack}, using the combination of the open-loop model and real-time voltage feedback. Next, we illustrate the results for these test cases.

\begin{figure}[h!]
    \centering
    \vspace{0mm}
    \includegraphics[scale = 0.55,trim={0 10mm 0 0mm}]{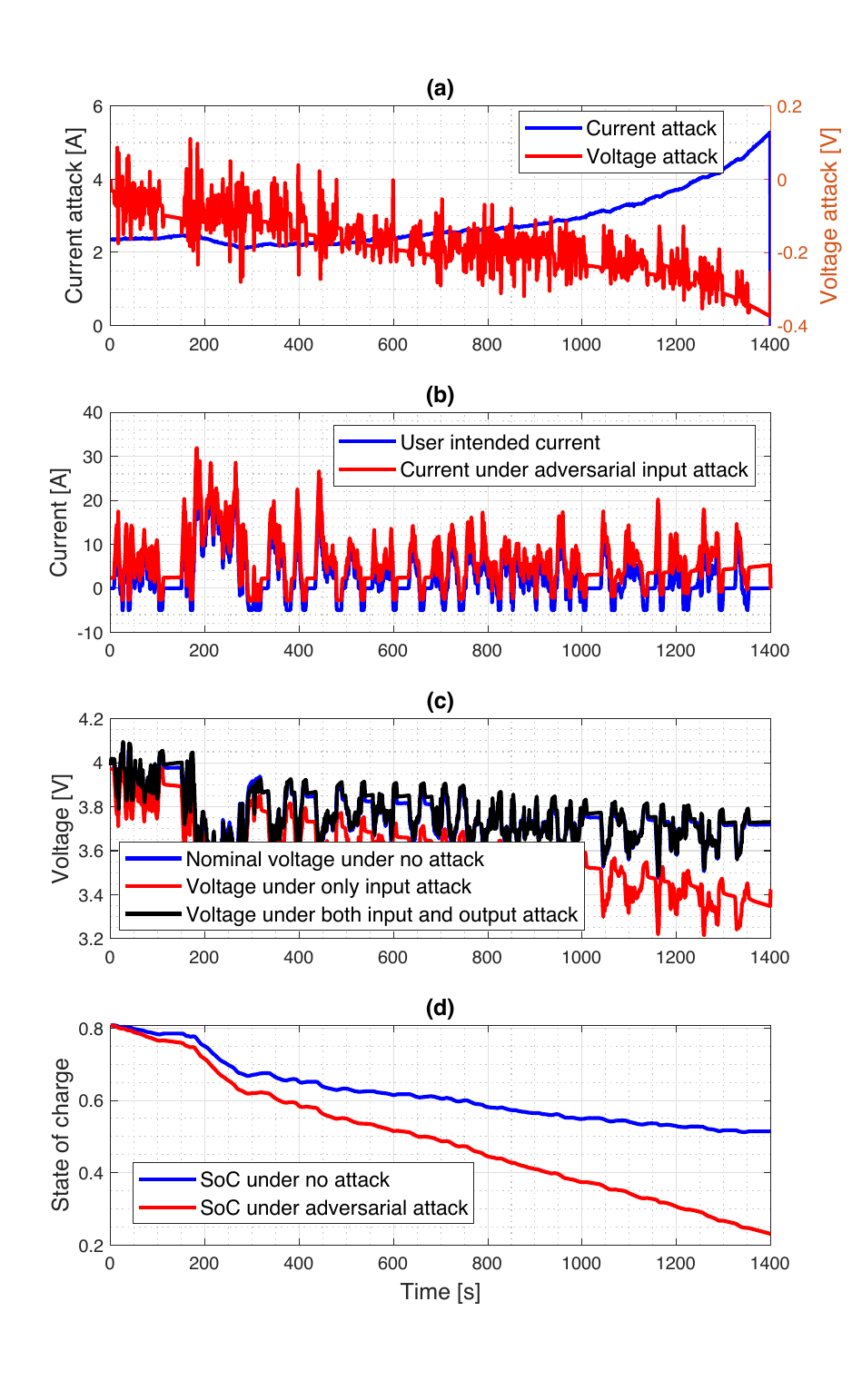}
    \vspace{0mm}
    \caption{Attack signals and resulting voltage, current, and State-of-Charge for Test Case 1. (a) Current and voltage attack signals; (b) user-intended current and adversarial current; (c) voltages under nominal scenario and various attacks; (d) State-of-Charge under attack and nominal conditions.}
    \label{fig:Results_UDDS_80_50}
\end{figure}

In Fig. \ref{fig:Results_UDDS_80_50}, the results for \textit{Test Case 1} are shown. In Fig. \ref{fig:Results_UDDS_80_50}(b), the ``User-intended current'' profile is shown. The generated input ``Current attack'' is shown in Fig. \ref{fig:Results_UDDS_80_50}(a), which serves the purpose of over-discharging the battery to 20\% $SoC$, ultimately leading to the applied ``Current under adversarial input attack'' shown in Fig. \ref{fig:Results_UDDS_80_50}(b). As a result, the battery is being over-discharged, as shown by the ``voltage under only input attack'' plot in Fig. \ref{fig:Results_UDDS_80_50}(c) and ``SoC under adversarial attack'' in Fig. \ref{fig:Results_UDDS_80_50}(d). Both of these plots indicate that the battery is being over-discharged. Next, to realize stealthiness, output ``Voltage attack'' is generated, as shown in Fig. \ref{fig:Results_UDDS_80_50}(a). This output voltage attack is added to the ``voltage under only input attack'' to generate ``voltage under both input and output attack''. As can be seen from Fig. \ref{fig:Results_UDDS_80_50}(c), the ``voltage under both input and output attack'' and ``Nominal voltage under no attack'' are very close to each other, thereby making the attack scenario almost indistinguishable from the no attack scenario. 

\begin{figure}[h!]
    \centering
    \vspace{0mm}
    \includegraphics[scale = 0.55,trim={0 10mm 0 0}]{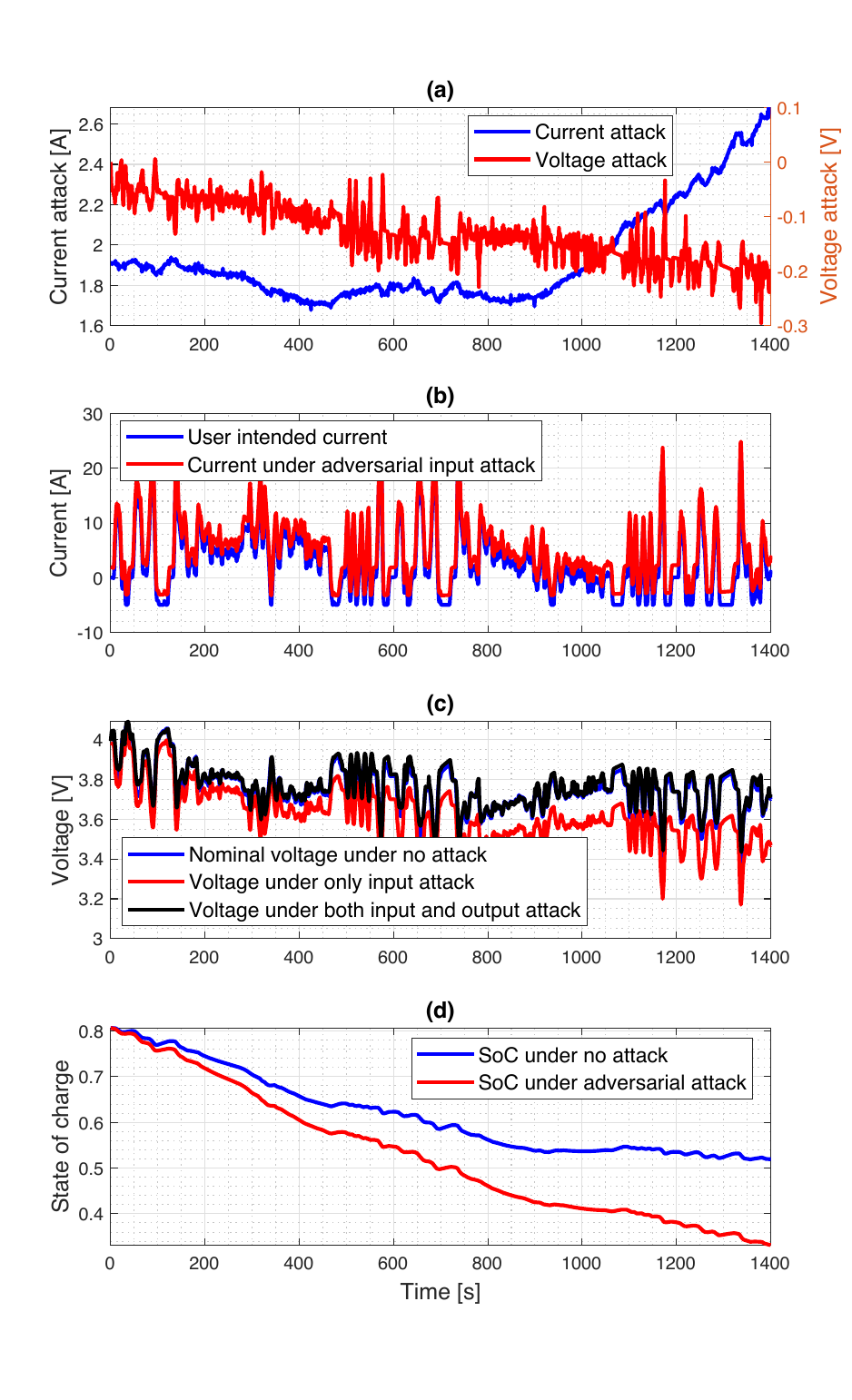}
    \vspace{0mm}
    \caption{Attack signals and resulting voltage, current, and State-of-Charge for Test Case 2. (a) Current and voltage attack signals; (b) user-intended current and adversarial current; (c) voltages under nominal scenario and various attacks; (d) State-of-Charge under attack and nominal conditions.}
    \label{fig:Results_US06_80_50}
\end{figure}

Similarly, the results for \textit{Test Case 2} are shown in Fig. \ref{fig:Results_US06_80_50} where the adversarial attack generation are explored for another over-discharging condition, but with a different dynamic current profile (derived from US06 driving schedule). The ``User-intended current'' profile is shown in Fig. \ref{fig:Results_US06_80_50}(b). Here the battery is over-discharged to 20\% $SoC$ using the generated input ``Current attack'' as shown in Fig. \ref{fig:Results_US06_80_50}(a). This ultimately leads to the applied current profile, ``Current under adversarial input attack'', as shown in Fig. \ref{fig:Results_US06_80_50}(b). The effect of this is, the battery gets over-discharged, as shown by the ``voltage under only input attack'' plot in Fig. \ref{fig:Results_US06_80_50}(c), and ``SoC under adversarial attack'' plot in Fig. \ref{fig:Results_US06_80_50}(d). Hence, both of these plots indicate that the battery is being over-discharged. Next, to achieve the objective of stealthiness, output ``Voltage attack'' is generated, as shown in Fig. \ref{fig:Results_US06_80_50}(a). To generate ``voltage under both input and output attack'', the output voltage attack mentioned earlier is added to the ``voltage under only input attack''. Fig. \ref{fig:Results_US06_80_50}(c) shows that the ``voltage under both input and output attack'' and ``Nominal voltage under no attack'' are very similar to each other, thereby making the attack scenario almost the same as the no attack scenario.

\begin{figure}[h!]
    \centering
    \vspace{0mm}
    \includegraphics[scale = 0.55,trim={0 10mm 0 0}]{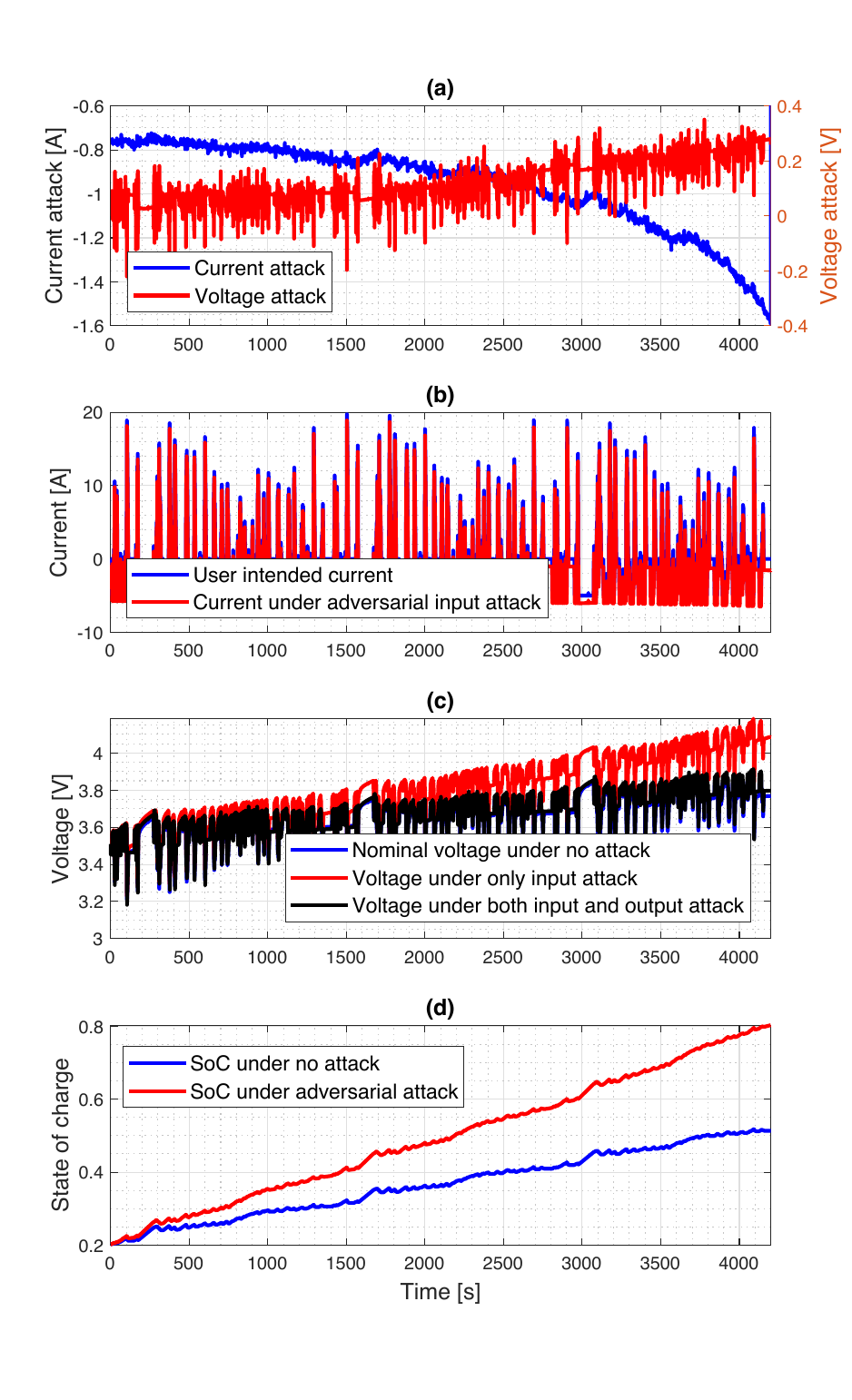}
    \vspace{0mm}
    \caption{Attack signals and resulting voltage, current, and State-of-Charge for Test Case 3. (a) Current and voltage attack signals; (b) user-intended current and adversarial current; (c) voltages under nominal scenario and various attacks; (d) State-of-Charge under attack and nominal conditions.}
    \label{fig:Results_UDDS_20_50}
\end{figure}

In Fig. \ref{fig:Results_UDDS_20_50}, the results for \textit{Test Case 3} are shown. In Fig. \ref{fig:Results_UDDS_20_50}(b), the ``User-intended current'' profile aims to charge the battery from 20\% to 50\% $SoC$. However, the adversary's intent is to overcharge to 80\% $SoC$. The generated input ``Current attack'' is shown in Fig. \ref{fig:Results_UDDS_20_50}(a), which serves the purpose of overcharging, ultimately leading to the applied ``Current under adversarial input attack'' shown in Fig. \ref{fig:Results_UDDS_20_50}(b). As a result, the battery is being overcharged, as shown by the ``voltage under only input attack'' plot in Fig. \ref{fig:Results_UDDS_20_50}(c) and ``SoC under adversarial attack'' in Fig. \ref{fig:Results_UDDS_20_50}(d). Both of these plots indicate that the battery is being overcharged. Next, to realize stealthiness, output ``Voltage attack'' is generated, as shown in Fig. \ref{fig:Results_UDDS_20_50}(a). This output voltage attack is added to the ``voltage under only input attack'' to generate ``voltage under both input and output attack''. As can be seen from Fig. \ref{fig:Results_UDDS_20_50}(c), the ``voltage under both input and output attack'' and ``Nominal voltage under no attack'' are very close to each other, thereby making the attack scenario almost indistinguishable from the no attack scenario. 

\begin{figure}[h!]
    \centering
    \vspace{0mm}
    \includegraphics[scale = 0.55,trim={0mm 20mm 0 0}]{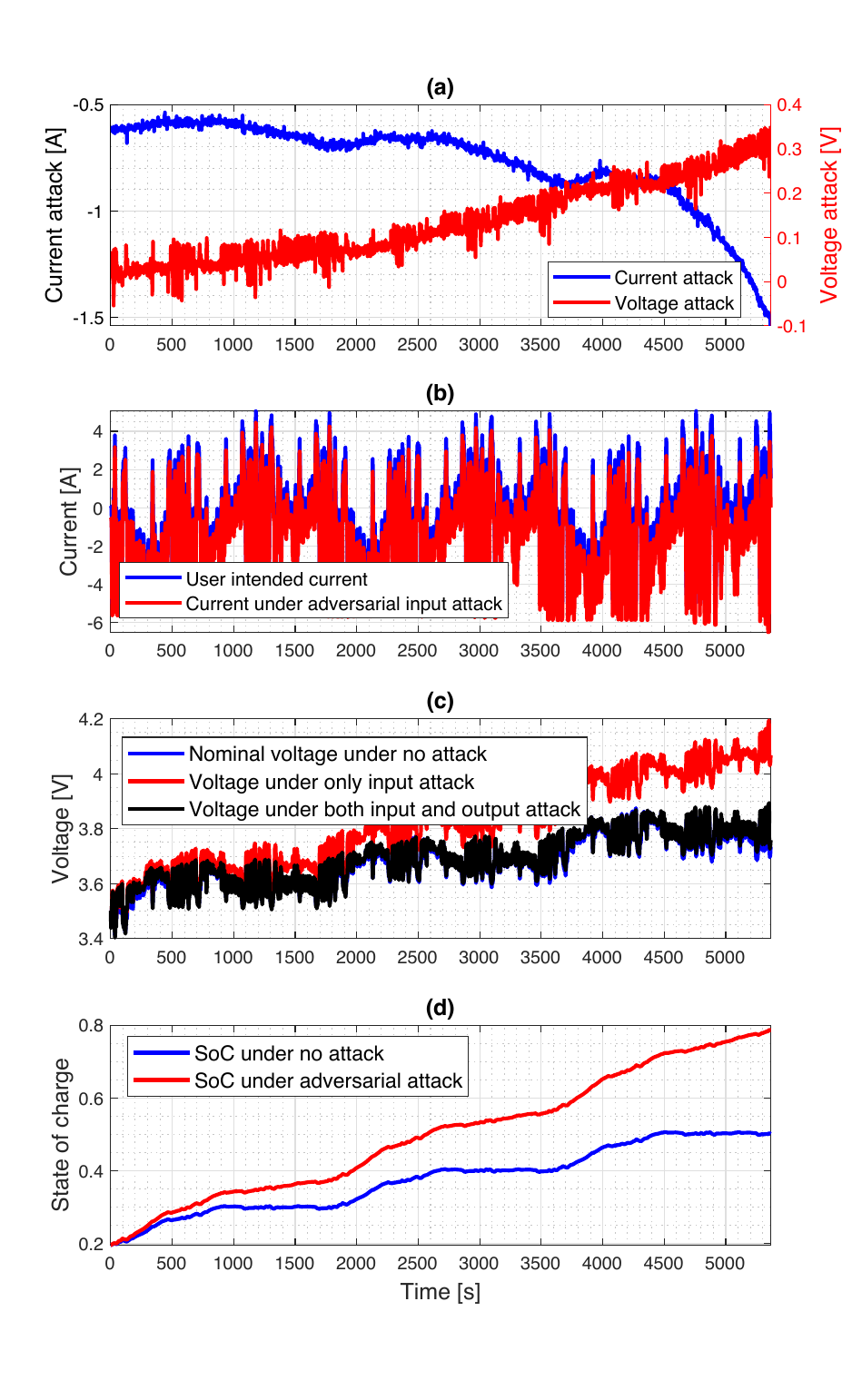}
    \vspace{0mm}
    \caption{Attack signals and resulting voltage, current, and State-of-Charge for Test Case 4. (a) Current and voltage attack signals; (b) user-intended current and adversarial current; (c) voltages under nominal scenario and various attacks; (d) State-of-Charge under attack and nominal conditions.}
    \label{fig:Results_US06_20_50}
\end{figure}

Similarly, the results for \textit{Test Case 4} are shown in Fig. \ref{fig:Results_US06_20_50} where the adversarial attack generation are explored for another overcharging condition, but with a different dynamic current profile (derived from US06 driving schedule). The ``User-intended current'' profile aims to charge the battery from 20\% to 50\% $SoC$ as shown in Fig. \ref{fig:Results_US06_20_50}(b). Here, the objective of the adversary is to overcharge to 80\% $SoC$. Here the battery is overcharged using the generated input ``Current attack'' as shown in Fig. \ref{fig:Results_US06_20_50}(a). This ultimately leads to the applied, ``Current under adversarial input attack'', as shown in Fig. \ref{fig:Results_US06_20_50}(b). The effect of this is, the battery gets overcharged, as shown by the ``voltage under only input attack'' plot in Fig. \ref{fig:Results_US06_20_50}(c), and ``SoC under adversarial attack'' in Fig. \ref{fig:Results_US06_20_50}(d). Hence, both of these plots indicate that the battery is being overcharged. Next, to achieve the objective of stealthiness, output ``Voltage attack'' is generated, as shown in Fig. \ref{fig:Results_US06_20_50}(a). To generate ``voltage under both input and output attack'', the output voltage attack mentioned earlier is added to the ``voltage under only input attack''. Fig. \ref{fig:Results_US06_20_50}(c) shows that the ``voltage under both input and output attack'' and ``Nominal voltage under no attack'' are very similar to each other, thereby making the attack scenario almost as same as the no attack scenario. 

\begin{figure}[h!]
    \centering
    \vspace{0mm}
    \includegraphics[scale = 0.6,trim={0mm 0mm 15mm 0mm}]{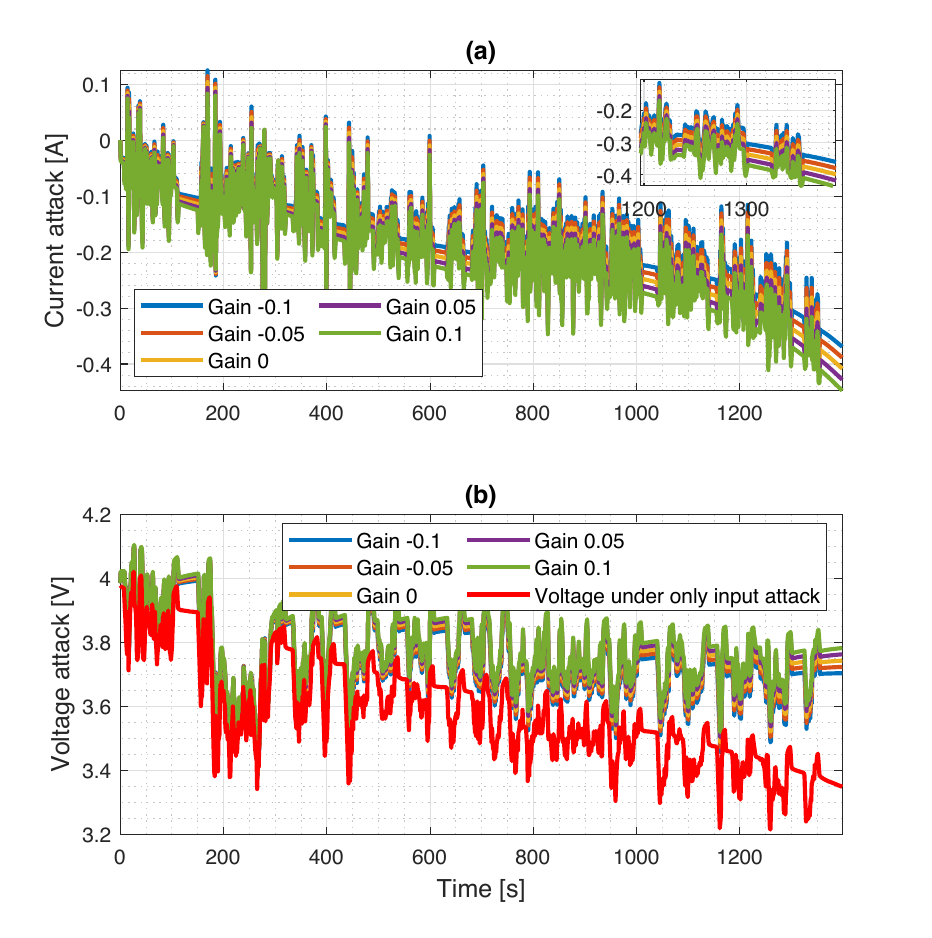}
    \vspace{0mm}
    \caption{{Sensitivity analysis of the output attack parameter for Test Case 1. (a) Voltage attack profiles for various values of the output attack gain parameter $K_a$. (b) Measured voltages under various values of the output attack gain parameter $K_a$.}}
    \label{fig:Sensitivity}
\end{figure}

{Next, we conduct a sensitivity study to understand the influence of adversarial output attack gain parameter $K_a$ (see \ref{eqn:output_attack2}) on the injected output attacks. Here, we use different values of $K_a$ to generate the voltage output attack profiles. Corresponding output attacks and measured voltages are shown in Fig. \ref{fig:Sensitivity}. The root mean square (RMS) errors between the nominal voltage (under no attack) and the voltage under attacks are given in Table \ref{table: error}. Note that the minimum RMS error is resulted for the gain value of $K_a=-0.05$. Hence, the attack gain parameter was chosen to be $-0.05$ in our test cases. 
}

\begin{table}[ht!]
\begin{center}
\caption{{Attack parameters ($K_a$) and corresponding RMS errors ($e_r$).}}
\label{table: error}
\small
\begin{tabular}{|c|c|c|c|c|c|} 
 \hline
 $K_a$ & $-0.1$ & $-0.05$ & $0$ & $0.05$ & $0.1$  \\ 
 \hline
 $e_r$ [V] & $0.0382$ & $0.038$ & $0.039$ & $0.04$ & $0.05$ \\
 \hline
\end{tabular}
\end{center}
\end{table}

\begin{figure}[h!]
    \centering
    \vspace{-2mm}
    \includegraphics[scale = 0.55]{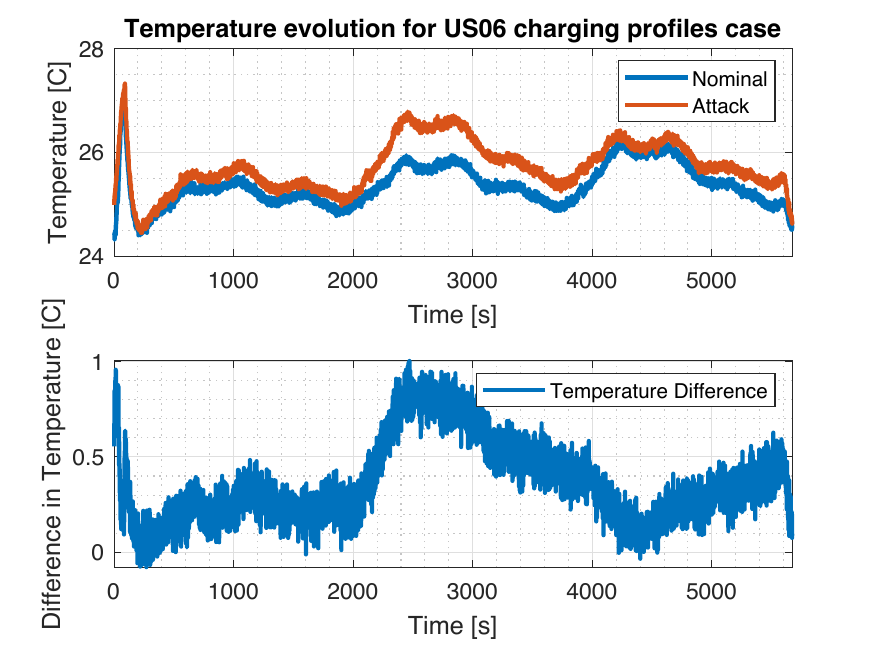}
    \vspace{-5mm}
    \caption{Comparison of temperature profiles in the presence and absence of attack, and their difference, under Test Case 4.}
    \label{fig:Temperature}
\end{figure}

In our next study, we have conducted additional experiments to explore the effects of attack on battery temperature. Specifically, we wanted to understand whether the stealthy attacks generated using the current formulation cause any significant change in temperature, which makes the attack signature significantly traceable on the temperature measurement. The result of this experiment is shown in Fig. \ref{fig:Temperature}. As can be seen in Fig. \ref{fig:Temperature}, we have found that the maximum difference between the temperature under attack and under no-attack scenarios is less than 1$^oC$. This 1$^oC$ change is not very significant for detecting this attack since such a change may also come from various factors, including external environment and cooling system variation.

\begin{figure}[h!]
    \centering
    \vspace{0mm}
    \includegraphics[scale = 0.55,trim={0 10mm 0 0}]{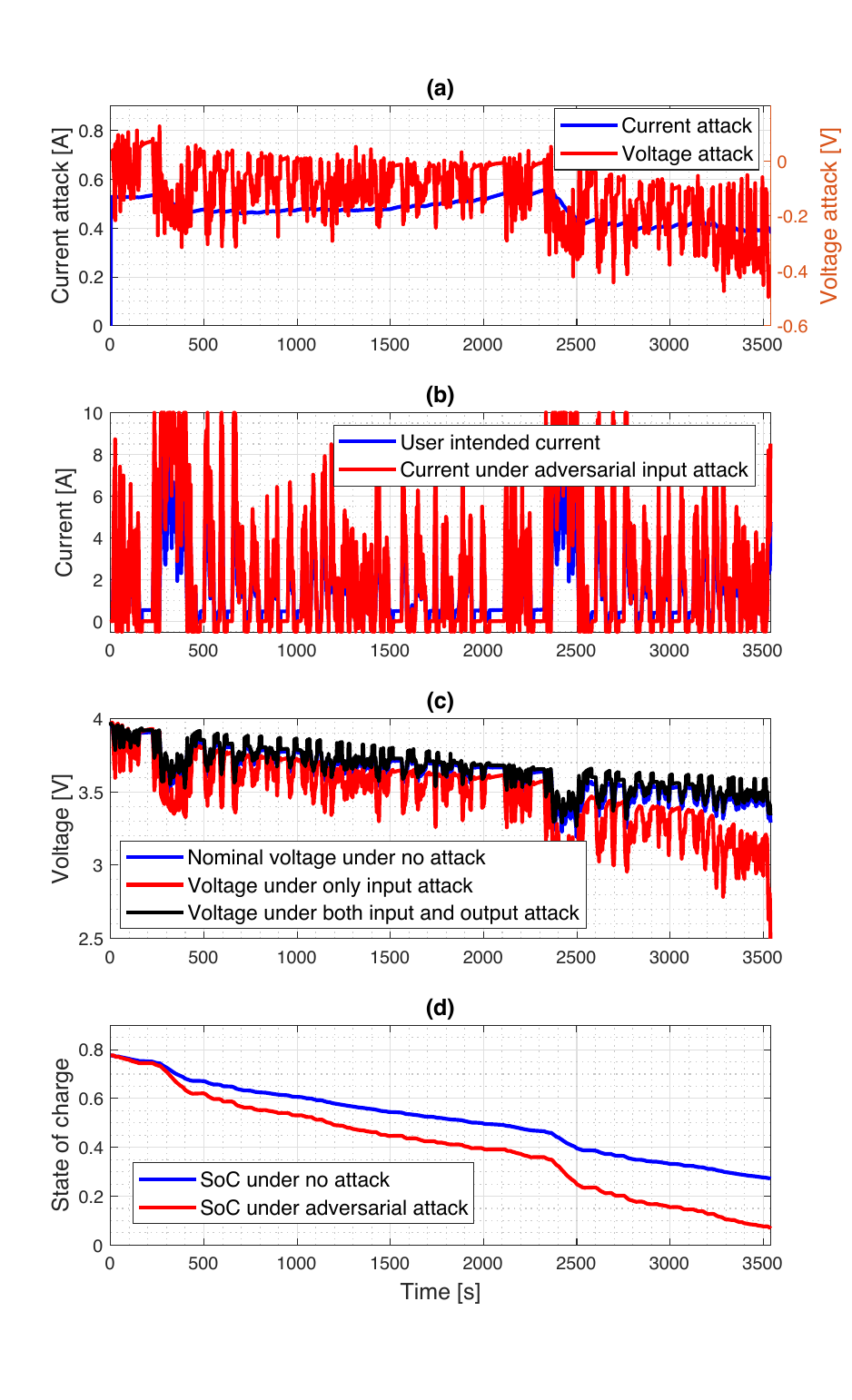}
    \vspace{0mm}
    \caption{Attack signals and resulting voltage, current, and State-of-Charge for Test Case 1 with  a different commercial battery cell. (a) Current and voltage attack signals; (b) user-intended current and adversarial current; (c) voltages under nominal scenario and various attacks; (d) State-of-Charge under attack and nominal conditions.}
    \label{fig:Results_UDDS_80_50_Panasonic}
\end{figure}

In our final study, we have conducted an additional experiment on a battery cell from a different manufacturer -- in order to understand the effectiveness of the proposed framework for different types of batteries. This new cell has the following parameters: Rated operating voltage range of 2.5V to 4.2V and rated capacity of 3085 mAh. The model parameters of this new cell have been identified, and subsequently, using the proposed approach, one attack case study was conducted. The corresponding results are presented in Fig. \ref{fig:Results_UDDS_80_50_Panasonic}. The results show that the proposed framework formulates the attack profiles which led to over-discharging, at the same time being stealthy. Hence, we can conclude that the proposed framework works on different types of batteries.

\section{Conclusion and future work}
The goal of this work is to create a form of threat model that can be used to design attack detection and mitigation strategies in battery systems. Here, optimal control principle is used to generate different input attack currents to overcharge or over-discharge the battery. After that, a combination of an open-loop model and real-time feedback is used to generate the output voltage attacks to potentially mask the effect of the input attack. These profiles are then tested on a commercial battery to evaluate their effectiveness. {As future extension of this foundational study, we plan to explore comprehensive active cyber-defense mechanisms for battery systems customized to various forms of adversarial behaviors. {We will also perform more comprehensive dynamical experimental studies to validate the proposed threat model.} Furthermore, we will also explore integration of threat models into the development of existing battery management technology. Along the lines of \cite{Srinath}, machine learning based paradigms for attack detection and mitigation in battery systems can also be explored as a complement to the model-based paradigms. Finally, expanding on our current work where only software-based vulnerabilities are considered, we also plan to explore adversarial threat models for batteries where hardware-based solutions are implemented.}

\ifCLASSOPTIONcaptionsoff
  \newpage
\fi

\bibliographystyle{IEEEtran}
\bibliography{ref}

\begin{thebibliography}{10}
\providecommand{\url}[1]{#1}
\csname url@samestyle\endcsname
\providecommand{\newblock}{\relax}
\providecommand{\bibinfo}[2]{#2}
\providecommand{\BIBentrySTDinterwordspacing}{\spaceskip=0pt\relax}
\providecommand{\BIBentryALTinterwordstretchfactor}{4}
\providecommand{\BIBentryALTinterwordspacing}{\spaceskip=\fontdimen2\font plus
\BIBentryALTinterwordstretchfactor\fontdimen3\font minus \fontdimen4\font\relax}
\providecommand{\BIBforeignlanguage}[2]{{%
\expandafter\ifx\csname l@#1\endcsname\relax
\typeout{** WARNING: IEEEtran.bst: No hyphenation pattern has been}%
\typeout{** loaded for the language `#1'. Using the pattern for}%
\typeout{** the default language instead.}%
\else
\language=\csname l@#1\endcsname
\fi
#2}}
\providecommand{\BIBdecl}{\relax}
\BIBdecl

\bibitem{gm_corporate_newsroom}
\BIBentryALTinterwordspacing
``General motors' future electric vehicles to debut industry's first wireless battery management system.'' [Online]. Available: \url{https://news.gm.com/newsroom.detail.html/Pages/news/us/en/2020/sep/ 0909-wbms.html}
\BIBentrySTDinterwordspacing

\bibitem{bosch_mobility_solutions}
\BIBentryALTinterwordspacing
``Battery in the cloud.'' [Online]. Available: \url{https://www.bosch-mobility-solutions.com/en/solutions/software-and-services/battery-in-the-cloud/battery-in-the-cloud/}
\BIBentrySTDinterwordspacing

\bibitem{bhusal2021cybersecurity}
N.~Bhusal, M.~Gautam, and M.~Benidris, ``Cybersecurity of electric vehicle smart charging management systems,'' in \emph{2020 52nd North American Power Symposium (NAPS)}.\hskip 1em plus 0.5em minus 0.4em\relax IEEE, 2021, pp. 1--6.

\bibitem{kharlamova2020cyber}
N.~Kharlamova, S.~Hashemi, and C.~Tr{\ae}holt, ``The cyber security of battery energy storage systems and adoption of data-driven methods,'' in \emph{2020 IEEE Third International Conference on Artificial Intelligence and Knowledge Engineering (AIKE)}.\hskip 1em plus 0.5em minus 0.4em\relax IEEE, 2020, pp. 188--192.

\bibitem{sripad2017vulnerabilities}
S.~Sripad, S.~Kulandaivel, V.~Pande, V.~Sekar, and V.~Viswanathan, ``Vulnerabilities of electric vehicle battery packs to cyberattacks,'' \emph{arXiv preprint arXiv:1711.04822}, 2017.

\bibitem{7408352}
I.~J. Martinez-Moyano, R.~Oliva, D.~Morrison, and D.~Sallach, ``Modeling adversarial dynamics,'' in \emph{2015 Winter Simulation Conference (WSC)}, 2015, pp. 2412--2423.

\bibitem{8813669}
A.~Khalid, A.~Sundararajan, A.~Hernandez, and A.~I. Sarwat, ``{FACTS Approach to Address Cybersecurity Issues in Electric Vehicle Battery Systems},'' in \emph{2019 IEEE Technology \& Engineering Management Conference (TEMSCON)}, 2019, pp. 1--6.

\bibitem{culler2021cybersecurity}
M.~Culler and H.~Burroughs, ``{Cybersecurity considerations for grid-connected batteries with hardware demonstrations},'' \emph{Energies}, vol.~14, no.~11, p. 3067, 2021.

\bibitem{10630530}
G.~Battista~Gaggero, A.~Armellin, G.~Ferro, M.~Robba, P.~Girdinio, and M.~Marchese, ``{BESS-Set: A Dataset for Cybersecurity Monitoring in a Battery Energy Storage System},'' \emph{IEEE Open Access Journal of Power and Energy}, vol.~11, pp. 362--372, 2024.

\bibitem{9507536}
M.~Pasetti, P.~Ferrari, P.~Bellagente, E.~Sisinni, A.~O. de~Sá, C.~B.~d. Prado, R.~P. David, and R.~C.~S. Machado, ``{Artificial Neural Network-Based Stealth Attack on Battery Energy Storage Systems},'' \emph{IEEE Transactions on Smart Grid}, vol.~12, no.~6, pp. 5310--5321, 2021.

\bibitem{7937798}
D.~D. Sharma, S.~N. Singh, J.~Lin, and E.~Foruzan, ``{Agent-Based Distributed Control Schemes for Distributed Energy Storage Systems Under Cyber Attacks},'' \emph{IEEE Journal on Emerging and Selected Topics in Circuits and Systems}, vol.~7, no.~2, pp. 307--318, 2017.

\bibitem{7723824}
X.~Liu, M.~Shahidehpour, Y.~Cao, L.~Wu, W.~Wei, and X.~Liu, ``{Microgrid Risk Analysis Considering the Impact of Cyber Attacks on Solar PV and ESS Control Systems},'' \emph{IEEE Transactions on Smart Grid}, vol.~8, no.~3, pp. 1330--1339, 2017.

\bibitem{rahman2018study}
S.~Rahman, H.~Aburub, Y.~Mekonnen, and A.~I. Sarwat, ``A study of ev bms cyber security based on neural network soc prediction,'' in \emph{2018 IEEE/PES Transmission and Distribution Conference and Exposition (T\&D)}.\hskip 1em plus 0.5em minus 0.4em\relax IEEE, 2018, pp. 1--5.

\bibitem{lee2021convolutional}
H.-J. Lee, K.-T. Kim, J.-H. Park, G.~Bere, J.~J. Ochoa, and T.~Kim, ``Convolutional neural network-based false battery data detection and classification for battery energy storage systems,'' \emph{IEEE Transactions on Energy Conversion}, vol.~36, no.~4, pp. 3108--3117, 2021.

\bibitem{dey2020cybersecurity}
S.~Dey and M.~Khanra, ``Cybersecurity of plug-in electric vehicles: Cyberattack detection during charging,'' \emph{IEEE Transactions on Industrial Electronics}, vol.~68, no.~1, pp. 478--487, 2020.

\bibitem{zografopoulos2021cyber}
I.~Zografopoulos, J.~Ospina, X.~Liu, and C.~Konstantinou, ``Cyber-physical energy systems security: Threat modeling, risk assessment, resources, metrics, and case studies,'' \emph{IEEE Access}, vol.~9, pp. 29\,775--29\,818, 2021.

\bibitem{bhattacharya2020automated}
A.~Bhattacharya, T.~Ramachandran, S.~Banik, C.~P. Dowling, and S.~D. Bopardikar, ``Automated adversary emulation for cyber-physical systems via reinforcement learning,'' in \emph{2020 IEEE International Conference on Intelligence and Security Informatics (ISI)}.\hskip 1em plus 0.5em minus 0.4em\relax IEEE, 2020, pp. 1--6.

\bibitem{khan2017stride}
R.~Khan, K.~McLaughlin, D.~Laverty, and S.~Sezer, ``Stride-based threat modeling for cyber-physical systems,'' in \emph{2017 IEEE PES Innovative Smart Grid Technologies Conference Europe (ISGT-Europe)}.\hskip 1em plus 0.5em minus 0.4em\relax IEEE, 2017, pp. 1--6.

\bibitem{Srinath}
\BIBentryALTinterwordspacing
S.~Srinath and S.~Dey, ``{A Machine Learning Approach toward Cyber-Physical Security of Battery Systems},'' \emph{Journal of Dynamic Systems, Measurement, and Control}, vol. 146, no.~6, p. 064503, 06 2024. [Online]. Available: \url{https://doi.org/10.1115/1.4065703}
\BIBentrySTDinterwordspacing

\bibitem{kim2018cloud}
T.~Kim, D.~Makwana, A.~Adhikaree, J.~S. Vagdoda, and Y.~Lee, ``Cloud-based battery condition monitoring and fault diagnosis platform for large-scale lithium-ion battery energy storage systems,'' \emph{Energies}, vol.~11, no.~1, p. 125, 2018.

\bibitem{trevizan2022cyberphysical}
R.~D. Trevizan, J.~Obert, V.~De~Angelis, T.~A. Nguyen, V.~S. Rao, and B.~R. Chalamala, ``Cyberphysical security of grid battery energy storage systems,'' \emph{IEEE Access}, vol.~10, pp. 59\,675--59\,722, 2022.

\bibitem{Kumbhar2018}
S.~Kumbhar, T.~Faika, D.~Makwana, T.~Kim, and Y.~Lee, ``{Cybersecurity for Battery Management Systems in Cyber-Physical Environments},'' in \emph{2018 IEEE Transportation Electrification Conference and Expo (ITEC)}, 2018, pp. 934--938.

\bibitem{KHARLAMOVA2023107795}
\BIBentryALTinterwordspacing
C.~T. Nina~Kharlamova and S.~Hashemi, ``{Cyberattack detection methods for battery energy storage systems},'' \emph{Journal of Energy Storage}, vol.~69, p. 107795, 2023. [Online]. Available: \url{https://www.sciencedirect.com/science/article/pii/S2352152X23011921}
\BIBentrySTDinterwordspacing

\bibitem{10230848}
V.~Obrien, V.~S. Rao, and R.~D. Trevizan, ``{Detection of False Data Injection Attacks in Battery Stacks Using Input Noise-Aware Nonlinear State Estimation and Cumulative Sum Algorithms},'' \emph{IEEE Transactions on Industry Applications}, vol.~59, no.~6, pp. 7800--7812, 2023.

\bibitem{9744036}
V.~Obrien, V.~Rao, and R.~D. Trevizan, ``{Detection of False Data Injection Attacks in Battery Stacks Using Physics-Based Modeling and Cumulative Sum Algorithm},'' in \emph{2022 IEEE Power and Energy Conference at Illinois (PECI)}, 2022, pp. 1--8.

\bibitem{barsoukov1999universal}
E.~Barsoukov, J.~H. Kim, C.~O. Yoon, and H.~Lee, ``Universal battery parameterization to yield a non-linear equivalent circuit valid for battery simulation at arbitrary load,'' \emph{Journal of power sources}, vol.~83, no. 1-2, pp. 61--70, 1999.

\bibitem{chen2020optimal}
Y.~Chen and X.~Zhu, ``Optimal attack against autoregressive models by manipulating the environment,'' in \emph{Proceedings of the AAAI Conference on Artificial Intelligence}, vol.~34, no.~04, 2020, pp. 3545--3552.

\bibitem{zhu2018optimal}
X.~Zhu, ``An optimal control view of adversarial machine learning,'' \emph{arXiv preprint arXiv:1811.04422}, 2018.

\bibitem{lessard2019optimal}
L.~Lessard, X.~Zhang, and X.~Zhu, ``An optimal control approach to sequential machine teaching,'' in \emph{The 22nd International Conference on Artificial Intelligence and Statistics}.\hskip 1em plus 0.5em minus 0.4em\relax PMLR, 2019, pp. 2495--2503.

\bibitem{lewis2012optimal}
F.~L. Lewis, D.~Vrabie, and V.~L. Syrmos, \emph{Optimal control}.\hskip 1em plus 0.5em minus 0.4em\relax John Wiley \& Sons, 2012.

\bibitem{bryson1975applied}
A.~Bryson and Y.-C. Ho, ``Applied optimal control, hemisphere,'' \emph{New York}, 1975.

\end{thebibliography}



\vspace{-45pt}
\begin{IEEEbiography}[{\includegraphics[width=1in,height=1.25in,clip,keepaspectratio]{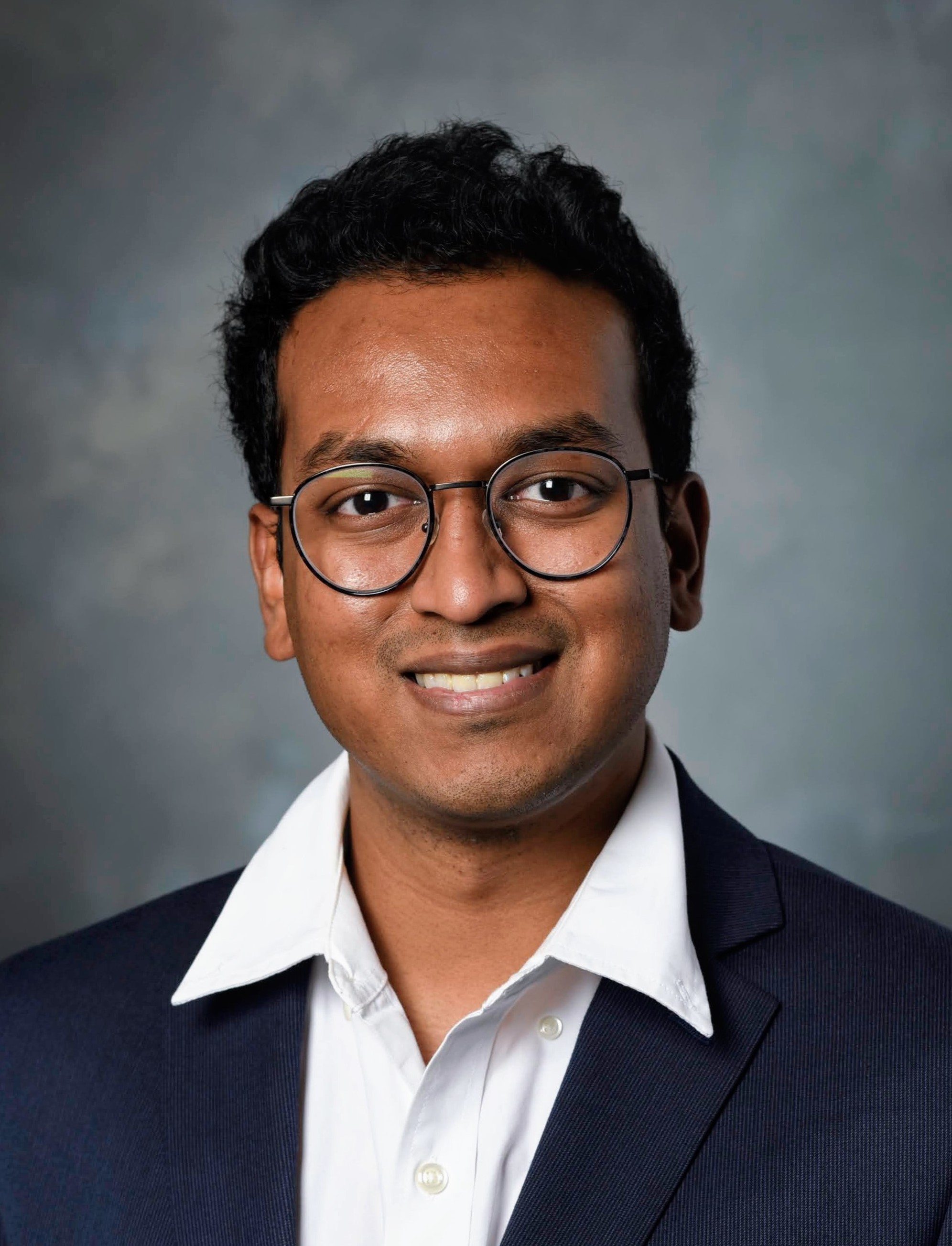}}]{Shanthan Kumar Padisala}
received B.E. in Manufacturing Engineering from BITS Pilani, India in 2019, followed by M.S. in Mechanical Engineering (Minor. Automotive) from The Ohio State University in 2021. Currently he is a Ph.D. candidate at The Pennsylvania State University working on Battery Systems. His interests involve reinforcement learning and optimal control applied in developing Battery Management Systems.
\end{IEEEbiography} 

\vspace{-43pt}

\begin{IEEEbiography}[{\includegraphics[width=1in,height=1.25in,clip,keepaspectratio]{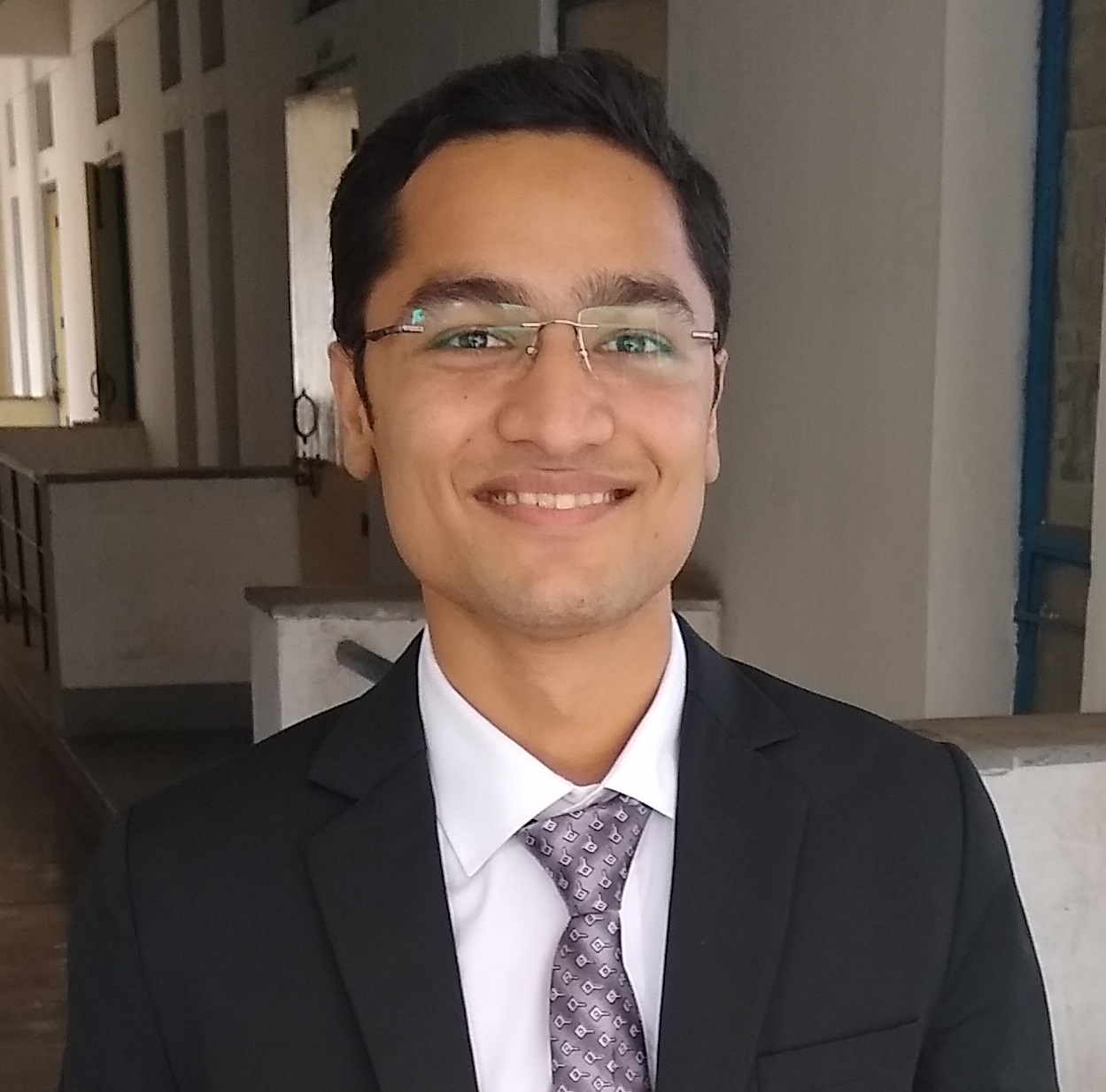}}]{Shashank Dhananjay Vyas}
received a dual B.Tech. + M.Tech in Mechanical Engineering from IIT Kharagpur, India in 2019. Later worked at Bajaj Automotive Ltd, India. Currently he is a Ph.D. candidate at The Pennsylvania State University working on Connected Autonomous Vehicles (CAVs). His primary research interests are optimisation, controls and machine learning in the realm of CAVs.
\end{IEEEbiography}

\vspace{-43pt}

\begin{IEEEbiography}[{\includegraphics[width=1in,height=1.25in,clip,keepaspectratio]{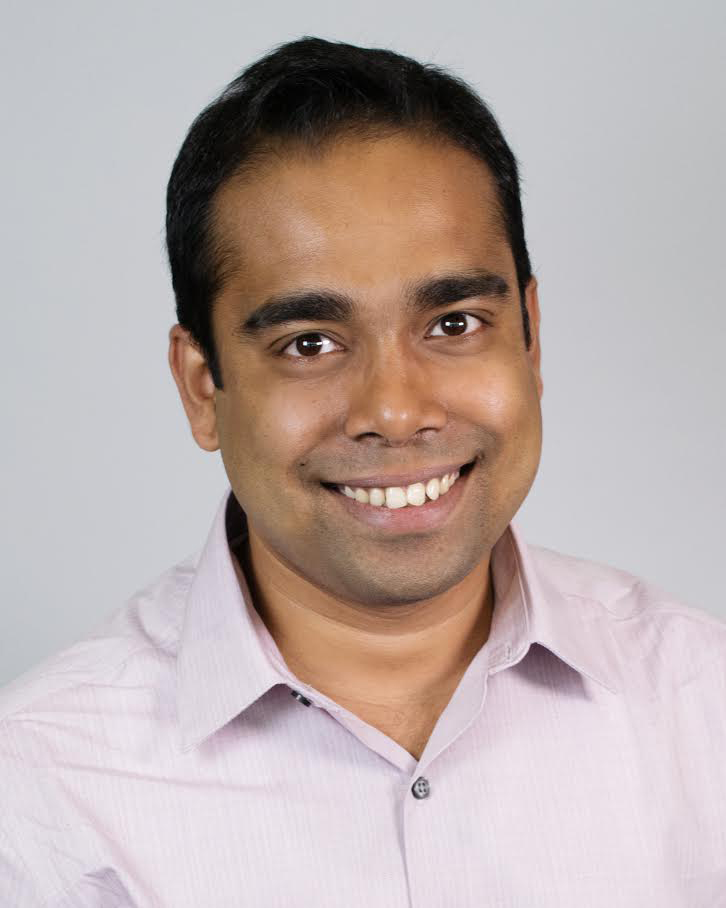}}]{Satadru Dey}
(Senior Member, IEEE) received the master’s degree in control systems from the Indian Institute of Technology Kharagpur, Kharagpur, India, in 2010, and the Ph.D. degree in automotive engineering from Clemson University, Clemson, SC, USA, in 2015.
He is an Assistant Professor with the Department of Mechanical Engineering, The Pennsylvania State University, University Park, PA, USA. From August 2017 to August 2020, he was an Assistant Professor with the University of Colorado Denver, Denver, CO, USA. He was a Postdoctoral Researcher with the University of California at Berkeley, Berkeley, CA, USA, from 2015 to 2017. His technical background is in the area of controls and his research interest lies in smart cities, energy, and transportation systems.
\end{IEEEbiography}
\end{document}